\documentclass{article}

\usepackage{arxiv}

\usepackage[utf8]{inputenc} 
\usepackage[T1]{fontenc}    
\usepackage{url}            
\usepackage{booktabs}       
\usepackage{amsfonts}       
\usepackage{nicefrac}       
\usepackage{microtype}      
\usepackage{lipsum}		
\usepackage{wrapfig}
\usepackage{graphicx}
\usepackage{multicol}
\usepackage{doi}

\usepackage{amsmath}
\usepackage{amssymb}
\usepackage{mathrsfs}
\usepackage{dsfont}
\usepackage{framed}

\usepackage{graphicx}
\graphicspath{ {./images/} }

\usepackage{graphicx}
\usepackage[table]{xcolor}

\usepackage{url}
\usepackage{cleveref}

\title{The Almost Sure Evolution of Hierarchy Among Similar Competitors}

\author{Christopher Cebra \\
	Department of Statistics\\
	University of Illinois Urbana-Champaign\\
	Champaign, IL  61820 \\
	 \\
	\And
	\href{https://orcid.org/0000-0001-6618-631X}{\includegraphics[scale=0.06]{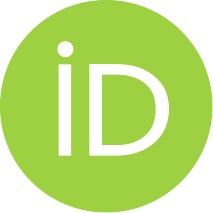}\hspace{1mm}Alexander Strang} \\
	Department of Statistics\\
	University of California-Berkeley\\
	Berkeley, CA 94720 \\
	\texttt{alexstrang@berkeley.edu} \\
}

\renewcommand{\shorttitle}{ Evolution of Hierarchy}

\hypersetup{
pdftitle={Evolution of Hierarchy},
pdfsubject={q-bio.NC, q-bio.PE},
pdfauthor={Christopher Cebra, Alexander Strang},
pdfkeywords={Competitive Systems, Evolutionary Game Theory, Transitivity, Cyclic Competition, Helmholtz-Hodge Decomposition},
}

\begin{document}
\maketitle

\begin{abstract}
	While generic competitive systems exhibit mixtures of hierarchy and cycles, real-world systems are predominantly hierarchical. We demonstrate and extend a mechanism for hierarchy; systems with similar agents approach perfect hierarchy in expectation. A variety of evolutionary mechanisms plausibly select for nearly homogeneous populations, however, extant work does not explicitly link selection dynamics to hierarchy formation via population concentration. Moreover, previous work lacked numerical demonstration. This paper contributes in four ways. First, populations that converge to perfect hierarchy in expectation converge to hierarchy in probability. Second, we analyze hierarchy formation in populations subject to the continuous replicator dynamic with diffusive exploration, linking population dynamics to emergent structure. Third, we show how to predict the degree of cyclicity sustained by concentrated populations at internal equilibria. This theory can differentiate learning rules and random payout models. Finally, we provide direct numerical evidence by simulating finite populations of agents subject to a modified Moran process with Gaussian exploration. As examples, we consider three bimatrix games and an ensemble of games with random payouts. Through this analysis, we explicitly link the temporal dynamics of a population undergoing selection to the development of hierarchy.
\end{abstract}


\maketitle


Hierarchies are widely observed in competitive systems. 
Nevertheless, hierarchies form a  small, highly organized, subset of the much broader space of possible competitive systems. Here mathematical theory and empirical observation diverge. Mathematically, hierarchy is a special case \cite{ralaivaosaona2020probability,Shizuka,strang2022network}. Empirically, hierarchy is often the rule \cite{Godoy,wright2016inhibitory,taylor1990allele,Gehrlein_b,van_Deemen_a}. Social choice provides a compelling example; voting cycles are possible, limit rational choice when they occur, and have been observed in real elections \cite{Flanagan,Gaubatz,Kurrild_a,Kurrild_c,Lagerspetz,Morse,Munkoe,Riker_a}, yet, are strikingly rare \cite{Regenwetter_a,Regenwetter_b,Regenwetter_d,van_Deemen_b}. 

The prevalence of hierarchy demands a mechanism. Domain-specific mechanisms include winner, loser, and bystander effects in animal behavior \cite{chase,Chase_b,drummond,dugatkin,hogeweg,Hsu,kawakatsu,Oliveira,Silk,vanhonk}, single-peaked preferences in social choice \cite{Black,Sen}, and ``spinning-top'' structures \cite{czarnecki2020real}. Domain-agnostic explanations also exist since the structure of a competing population depends both on the possible advantage relations and on the realized population of agents. Many populations evolve in response to their advantage relations, so steady states may exhibit competitive structures that would otherwise be considered special cases. Dynamic or not, the choice of population may suppress cycles \cite{park2020cyclic}. For example, Go is traditionally considered a highly transitive game, yet recent work in adversarial policies has demonstrated the existence of agents who defeat superhuman Go A.I.'s, but lose to amateur human players \cite{wang2023adversarial}. Similar work discovered cycles in Go by training for diversity in addition to performance \cite{omidshafiei2020navigating}.

Simple population structures may limit cyclicity. Indeed, recent work has shown that increasing agents' similarity suppresses cyclicity and promotes hierarchy when the underlying payout function is smooth \cite{cebra2023similarity}. Evolutionary game theory \cite{Maynard_Smith,hofbauer2003folk} provides a variety of canonical scenarios in which an evolving population concentrates \cite{Cressman_c,cabrales2000stochastic,Dieckmann,foster1990stochastic,fudenberg1992evolutionary,imhof2005long,kandori1993learning,oechssler2001evolutionary}. In a concentration limit, sampled populations approach perfect hierarchy provided the self-play gradient is non-zero where the population concentrates. The latter condition separates populations that converge at a boundary of the space of possible agents from populations that converge at interior equilibria. Only the latter support cycles. Moreover, the smoother the underlying payout, the less concentration is required for hierarchy. These results provide a firm mathematical grounding for analogous observations in simulated evolution experiments \cite{park2020cyclic}.

We extend these result by exchanging a stronger notion of convergence for a weaker, more widely used, notion of hierarchy. We then show how to analyze hierarchy formation under different learning rules. In Section \ref{sec: replicator dynamic}, we provide a dynamical analysis for populations subject to the continuous replicator dynamic with diffusion. In Section \ref{sec: steady state cyclicity}, we use the theory to compare learning rules. Finally, in Section \ref{sec: numerics}, we demonstrate the concentration mechanism numerically.

\section{Theory} \label{sec: theory}

\subsection{Introduction} \label{sec: intro theory}

Let $(f,\Omega)$ define a functional form game, where $f:\Omega \times \Omega \rightarrow \mathbb{R}$ is a zero-sum performance function that accepts the traits of two agents, $x,y \in \Omega$, and returns a measure of the advantage $x$ possesses over $y$, $f(x,y)$. Performance could denote expected payout, utility, log odds, or some other measure of advantage such that $f(x,y) = -f(y,x)$. The trait space $\Omega$ is similarly general, but we will usually consider $\Omega \subseteq \mathbb{R}^T$ where $T$ is the number of relevant traits. Traits could represent physical characteristics, probabilities of action, parameters in a policy function, location in a latent behavior space, or even hyperparameters for training.

Empirical game theory studies the restriction of a functional form game to a competitive network $\mathcal{G} = (\mathcal{V},\mathcal{E})$, with observed pairwise relations, $\mathcal{E}$, sampled between a set of $V = |\mathcal{V}|$ agents, $\mathcal{V}$, with traits $\mathcal{X} = \{x_1,x_2,\hdots,x_{V}\}$. Let $\mathcal{G}_{\rightarrow}$ denote the directed version of $\mathcal{G}$ with edges pointing from $i$ to $j$ for all ordered pairs such that $f(x_j,x_i) \geq 0$. That is, for all pairs such that $j$ possesses an advantage over $i$, denoted $j \succeq i$. 

A competitive network $\mathcal{G}$ is \textit{transitive} if the directed network $\mathcal{G}_{\rightarrow}$ is acyclic, i.e.~there are no advantage cycles $\mathcal{C}=\{i_1,i_2,\hdots, i_n, i_{n+1} = i_1 \}$ such that $i_{m+1} \succeq i_{m}$ for all $m$. Thus, transitive networks are hierarchies. A network is \textit{intransitive} if $\mathcal{G}_{\rightarrow}$ contains an advantage cycle. 

Transitivity and intransitivity only depend on the direction of advantage, i.e.~the sign  $f$ restricted to the sampled pairs. These definitions fail to distinguish between, or make any prediction regarding, the magnitude of advantage, so treat minor and major advantages in the same way. Moreover, by reducing hierarchy to a condition on the topology of $\mathcal{G}_{\rightarrow}$, the traditional definitions introduce unnecessary computational problems. For example, Slater's measure \cite{Slater}, the minimum number of edge reversals (upsets) needed to convert to a transitive system, is NP-hard to compute \cite{Charbit,Endriss}.

Stronger notions follow. Strong transitivity requires that $f(x_{i(1)},x_{i(m)})$ is nondecreasing in $m$ for any sequence such that $x_{i(m+1)} \succeq x_{i(m)}$. Perfect transitivity enforces yet stronger conditions; there exists some rating function $r:\Omega \rightarrow \mathbb{R}$, such that $f(x_i,x_j) = r(x_i) - r(x_j)$ for all sampled pairs $(i,j) \in \mathcal{E}$. This is the standard hypothesis used by most rating systems, such as Elo or Bradley-Terry \cite{Aldous,Hvattum,Kovalchik,Langville}. A competitive system is perfectly transitive if and only if it is curl-free, that is, the sum of $f$ around any cycle equals zero. Similarly, a competitive system is perfectly cyclic if it is favorite-free, i.e.~no competitor $i$ has an average advantage over all of it's neighbors \cite{strang2022network}. While perfect transitivity and cyclicity are special cases, all competitive networks admit a unique decomposition into perfectly transitive and cyclic components. Moreover, the decomposition can be performed at scale by solving a sparse least squares problem for $r$ \cite{strang2022network}. The necessary decomposition is the Helmholtz-Hodge Decomposition (HHD), familiar in physics, simplicial homology, and network flows \cite{candogan2011flows,jiang2011statistical,lim2020hodge,strang2020applications}.

The realized structure of a competitive system depends on the underlying population. Let $\pi_x$ denote a distribution of traits, and $F$ denote the set of observed advantage relations between a population of agents with traits drawn i.i.d.~from $\pi_x$. Let, $\rho$ denote the correlation between $f(X,Y)$ and $f(X,W)$ for $X,Y,W$  drawn i.i.d from $\pi_x$. The correlation $\rho$ can take any value between 0 and 0.5. If $\rho = 0.5$, the system is perfectly transitive so all advantage relations reduce to differences in agent ratings \cite{strang2022network}. 
For $\rho < 0.5$, populations exhibit a mixture of transitive and cyclic components, $F_t$ and $F_c$. The degree of hierarchy in a system can be evaluated by comparing their sizes \cite{jiang2011statistical,strang2022network}. The expected degree of hierarchy in a system increases monotonically as $\rho$ approaches 0.5, so the difference between $\rho$ and 0.5 controls convergence to hierarchy in expectation \cite{strang2022network}.

In \cite{cebra2023similarity} the authors introduce a ratio of moments, $\epsilon$ such that $\rho \leq (2 (1 + \epsilon))^{-1}.$ The expected size of the cyclic and transitive components are $\mathcal{O}(\epsilon)$ and $\mathcal{O}(1)$ in $\epsilon$, so, the smaller $\epsilon$, the more transitive the expected system.

If $f$ is linear, then performance is perfectly transitive. If $f(x,y)$ is nearly linear in the neighborhood supporting the population, then $\epsilon$ is small, so $\rho$ is near 0.5. More generally, if $f$ admits a non-vanishing linearization about $z$, converges to its linearization on small neighborhoods, and $\pi_x$ is sufficiently concentrated about $z$, then $\epsilon$ can be made arbitrarily small. It follows that populations of similar agents are highly transitive in expectation. Sufficient assumptions for convergence to perfect transitivity and the associated rates are studied in \cite{cebra2023similarity}. Those results are summarized, then strengthened here through extended analysis, example, and demonstration.

\subsection{Convergence to Transitivity in Expectation} \label{sec: convergence}

For $x,y$ near $z$, write:
\begin{equation}
    f(x,y) = r(x|z) - r(y|z) + h(x,y|z)
\end{equation}
where $r(\cdot|z)$ is a linear local rating function that linearizes $f$ about $z$ and $h(x,y|z)$ denotes higher-order terms. To study convergence rates in expectation, first assume that $h$ is, at most, quadratic. Then, following \cite{cebra2023similarity}:

\textbf{Lemma 1: [Quadratic Performance and $\epsilon$]} \label{Lem: quadratic performance}
\textit{If $f$  is quadratic, then:
\begin{equation} \label{eqn: quadratic epsilon}
\rho = \frac{1}{2} \frac{1}{1 + \epsilon}, \textit{ and } 
    \epsilon = \frac{\mathbb{E}_{X,Y}[h(X,Y|z)^2]}{2 \mathbb{V}_{X}[r(X|z)]}.
\end{equation}
so, by the trait-performance theorem \cite{strang2022network}, $\mathbb{E} \left[\|F\|^2 \right] = \mathcal{O}(1)$, $\mathbb{E} \left[\|F_t\|^2 \right] = \mathcal{O}(1)$, and $\mathbb{E} \left[\|F_c\|^2 \right] = \mathcal{O}(\epsilon)$ in $\epsilon$.}

Generic $f$ are not quadratic, but when sufficiently smooth, $\epsilon$ can be approximated by the quadratic approximation of the full performance function \cite{cebra2023similarity}. 

The quadratic approximation to $f$ about $z$ sets:
\begin{equation}
\begin{aligned}
    & h(x,y|z) \simeq r_2(x|z) - r_2(y|z) + (x - z)^{\intercal} H_{xy}(z) (y - z) \\
    & r_2(x|z) = r(x|z) + \frac{1}{2}(x - z)^{\intercal} H_{xx}(z) (x - z)
    \end{aligned}
\end{equation}
where $r_2$ is the quadratic local rating function, $H(z)$ is the Hessian of $f(x,y)$ evaluated at $x = y = z$, the subscript $xx$ denotes mixed partials concerning the traits of the first agent, and $xy$ denotes mixed partials concerning the traits of the first and second agents. The blocks are symmetric and skew-symmetric respectively. To second-order, all intransitivity is generated by $H_{xy}$. 

The local quadratic model correctly predicts $\epsilon$ in a concentration limit when $f$ is sufficiently smooth. We will say that $f$ is quadratically approximable at $z$ if $f(x,y)$ is second differentiable at $x = y = z$ and its second-order Taylor expansion about $x = y = z$ has errors that are bounded by a cubic polynomial of $(x - z)$ and $(y-z)$ on some ball centered at $z$ with nonzero radius. The latter holds whenever $f$ is analytic, or has Lipschitz continuous second derivatives.

\textbf{Theorem 1: [Trait Concentration]}
\textit{ Suppose $f(x,y)$ is a bounded performance function on $\Omega \times \Omega$ and is quadratically approximable on $z$. Suppose that $\pi_x$ has centroid $z$ and depends on a concentration parameter $\kappa$ such that the probability $p(\kappa)$ of sampling $X$ outside a ball $B_{R(\kappa)}(z)$ goes to zero  at rate $\mathcal{O}_{<}(\kappa^4)$ for some $R(\kappa)$ that converges to zero at rate $\mathcal{O}_{<}(\kappa^{4/5})$. Then:}

\begin{enumerate}

\item \textit{$\epsilon$ converges to its approximations using the local quadratic model of $f(x,y)$ about $x = y = z$ with errors vanishing faster than the approximations terms.} 

\item \textit{If $\nabla_x f(x,y)|_{x=y=z} \neq 0$:}
$$    
\begin{aligned}
    \mathbb{E}[||F_t||^2] = \mathcal{O}_{=}(\kappa^2), \quad \mathbb{E}[||F_c||^2] = \mathcal{O}_{\leq}(\kappa^4)
\end{aligned}
$$
\textit{with equality if and only if $H_{xy}(z,z) \neq 0$.}

\item \textit{If $\nabla_x f(x,y)|_{x=y=z} = 0$  and $H_{xx}(z,z) \neq 0$ then:}
$$
\begin{aligned}
    \mathbb{E}[||F_t||^2] = \mathcal{O}_{=}(\kappa^4), \quad \mathbb{E}[||F_c||^2] = \mathcal{O}_{\leq}(\kappa^4) 
\end{aligned}
$$
\textit{with equality if and only if $H_{xy}(z,z) \neq 0$. }
\end{enumerate}

Here $\mathcal{O}_{=}(h(\kappa))$, $\mathcal{O}_{\leq}(h(\kappa))$, and $\mathcal{O}_{<}(h(\kappa))$ denote convergence to zero at a rate equal to, equal to or faster than, and faster than $h(\kappa)$. Table \ref{tab: rate of convergence} summarizes the concentration theory.

\begin{table}[h]
\begin{centering}
\begin{tabular}{ |p{2.5cm}| p{2.1cm}|p{6cm}|  }
 \hline
 \multicolumn{3}{|c|}{Predicted Rates of Convergence} \\
 \hline
 Scenario & Convergence Rate & Plausible Scenario\\
 \hline
 $g\neq0$, $H_{xy} \neq 0$ &  Quadratic   & Multidimensional $\Omega$, centroid at boundary NE or away from interior NE\\
 $g \neq 0$, $H_{xy} = 0$ & Quartic & One-dimensional $\Omega$, centroid at boundary NE or away from interior NE\\
 $g = 0$, $H_{xy} \neq 0$ & Sustains cycles  & Multi-dimensional $\Omega$, centroid at interior NE \\
 $g = 0 $, $H_{xy} = 0$   &  Quadratic   & One-dimensional $\Omega$, centroid at interior NE\\
 \hline
\end{tabular}

 \end{centering}
 \vspace{2mm}
 \caption{Summary of predicted convergence rates from Theorem 1 applied to example scenarios classified by the lowest order derivatives. Note, in all cases we assume $H_{xx} \neq 0$.}
\label{tab: rate of convergence}
\end{table}

\vspace*{-\baselineskip}

\subsection{Almost Sure Transitivity} \label{sec: almost sure transitivity}

Theorem 1 shows that, if the population concentrates where the local linearization is non-zero, then the competitive network converges in expectation to perfect transitivity. Here, we weaken our notion of transitivity to strengthen our notion of convergence. Instead of proving convergence to perfect transitivity in expectation, we prove convergence to transitivity in probability. We outline the proof below and provide details in the appendix.

First, adding new edges to a network can only convert a transitive network to an intransitive network by creating a cycle. Therefore, if the completion of $\mathcal{G}$ is transitive, so is $\mathcal{G}$. Once completed, transitivity is determined by the set of triangles. When traits are drawn i.i.d., the triangles are exchangeable, so the probability that a single triangle is transitive bounds the probability that the entire network is transitive. A triangle is transitive if the cyclic component is sufficiently small relative to the transitive component \cite{strang2022network}: 

\textbf{Lemma 2 [Transitivity of Triangular Networks]} \textit{A triangular competitive network is transitive if $\|f_c\|^2 \leq \frac{1}{3} \|f\|^2$.}

Accordingly, we can bound the probability an arbitrary network is transitive from below with the probability that $\|F_c\|/\|F\|$ is sufficiently small on a single triangle.

To separate $\|F_c\|$ and $\|F\|$, we seek a function $\alpha(\kappa)$ that acts as an upper bound on $\|F_c\|^2$, a lower bound on $\frac{1}{3}\|F\|^2$, and converges to zero as function of a concentration parameter $\kappa$. If such a bound holds, then the rate of convergence of $\alpha(\kappa)$ to zero determines whether the network converges to transitivity in probability. Specifically, if $\alpha(\kappa)$ converges to zero slower than $\mathbb{E}[\|F_c\|^2]$, but faster than $\mathbb{E}[\|F\|^2]$, then  $\|F_c\|^2 \leq \alpha(\kappa) \leq \frac{1}{3} \|F\|^2$ will hold with high probability.

\textbf{Theorem 2: } \textit{Suppose that the smoothness and concentration assumptions of Theorem 1 hold, and $\nabla_x f(x,y)|_{x=y=z} \neq 0$. If, in addition, there exists an $\alpha(\kappa) = \mathcal{O}_{>}(\kappa^4)$ such that:}
\begin{equation}
    \text{Pr}\{f(Y,W)^2 > 3 \alpha(\kappa) \} \rightarrow 1 
\end{equation}
\textit{for, trait vectors $Y,W$ drawn i.i.d.~from $\pi_x(\kappa)$ then:}
\begin{equation}
    \text{Pr}\{\mathcal{G}(X,f,\mathcal{E}) \text{ is transitive}\} \rightarrow 1
\end{equation}
\textit{as $\kappa \rightarrow 0$.}

It remains to show that, under reasonable concentration assumptions, there exists such an $\alpha(\kappa)$. Note that, as the population concentrates, $f(Y,W)^2$ converges to zero in expectation since $f$ is smooth and $f(x,x) = 0$ for all $x$. 
To ensure that $\|F\|$ does not vanish too quickly, we need to bound the concentrating distribution from below.

\textbf{Definition 1: } A sequence of distributions $\pi_x(\kappa)$ \textit{concentrates to $z$ at rate $\mathcal{O}_{=}(h(\kappa))$} if $h(\kappa) \rightarrow 0$ as $\kappa \rightarrow 0$, and:
\begin{enumerate}
    \item \textit{for any $R(\kappa) = \mathcal{O}_{>}(h(\kappa))$, $\text{Pr}\{|X - z| > R(\kappa)\} \rightarrow 0$ for $X \sim \pi_x(\kappa)$, and}
    \item \textit{for any sequence of nested sets $S(\kappa) \subseteq \Omega \subseteq \mathbb{R}^T$, $S(\kappa') \subseteq S(\kappa)$ if $\kappa' \leq \kappa$ , whose volume $\text{Vol}(S(\kappa))$ is $\mathcal{O}_{<}(h(\kappa)^T)$, $\text{Pr}\{X \in S(\kappa)\} \rightarrow 0$ as $\kappa \rightarrow 0$.} 
\end{enumerate}

If $f$ is quadratically approximable and $\pi_x(\kappa)$ converges at rate $\mathcal{O}_{=}(\kappa)$ then the desired $\alpha(\kappa)$ exists.

\textbf{Theorem 3: [Almost Sure Transitivity]} 
\textit{Assume that the conditions of Theorem 1 hold, and $\pi_x(\kappa)$ concentrates to $z$ at rate $\mathcal{O}_{=}(\kappa)$. Then, if $\nabla_x f(x,y)|_{x=y=z} \neq 0$:}
\begin{equation}
    \lim_{\kappa \rightarrow 0} \text{Pr}\{\mathcal{G}(X,f,\mathcal{E} \text{ is transitive}\} = 1.
\end{equation}


\vspace{2mm}

\subsection{Example Selection Dynamics} \label{sec: selection dynamics}
The replicator equation is widely used to model the evolution of a population where payout translates to per capita growth rate. For any strategy $x$, the per capita rate of change in the proportion playing strategy $x$ equals the difference between the average payout to $x$ and the average payout of all available strategies. The replicator equation is typically derived as the mean-field, or infinite population, limit of a discrete population process \cite{schuster}. Popular agent-based methods like imitation learning or the Moran process \cite{binmore,borgers1997learning} can approximate the replicator dynamic.

Steady-state distributions provide dynamic justification for static equilibrium concepts from game theory. For example, an Evolutionarily Stable State (ESS) is a strategy that cannot be invaded by a single mutant strategy \cite{Maynard_Smith}. An ESS on the interior of the strategy space is a globally asymptotically stable rest point for the replicator dynamic, while an ESS on the boundary is locally asymptotically stable  \cite{hofbauer2003folk}. Convergence towards a monomorphic population at an ESS, or other equilibrium, provides a natural concentration mechanism.

To incorporate a mutation/exploration term into the replicator equation, we add a diffusive term with diffusivity $D$:
\begin{equation} \label{eqn: continuous rep}
    \partial_t \pi(x,t) =  \mathbb{E}_{Y \sim \pi(t)}[f(x,Y)] \pi(x,t) + \nabla \cdot D(x) \nabla \pi(x,t).
\end{equation}

For concentration analysis, we focus on the quadratic case:
\begin{equation} \label{eqn: f quad}
\begin{aligned}
    & f(x,y) = r(x) - r(y) + x^{\intercal} H_{xy} y \\
    & r(x) = g^{\intercal} x + \frac{1}{2} x^{\intercal} H_{xx} x
\end{aligned}
\end{equation} 
where $g$ is the gradient $\nabla_{x} f(x,y)|_{x=y=0}$.

When $f$ is quadratic, Gaussian distributions are invariant under the replicator dynamic \cite{cressman2004coevolution}. The same holds after incorporating diffusion:

\textbf{Lemma 3: [Gaussian Invariance]} \textit{If $f$ is quadratic, then the manifold of multivariate normal distributions is invariant under the continuous replicator dynamic with diffusion \ref{eqn: continuous rep}. Moreover, subject to \ref{eqn: continuous rep}, and given quadratic $f$ of the form \ref{eqn: f quad}, the mean $\bar{x}(t) = \mathbb{E}_{X\sim \pi(t)}[X]$ and covariance $\Sigma(t) = \text{Cov}_{X\sim \pi(t)}[X]$ are governed by the coupled system of ODE's:
\begin{equation} \label{eqn: moment dynamics}
    \begin{aligned}
    &\frac{d}{dt}\bar{x}(t) =  \Sigma(t) (g + J \bar{x}(t))\\
    &\frac{d}{dt} \Sigma(t) = D + \Sigma(t)^{\intercal} H_{xx} \Sigma(t) 
\end{aligned}
\end{equation}
where $J = H_{xx} + H_{xy}$ is the Jacobian of the optimal self-training vector field $\nabla_{x} f(x,y)|_{x=y=z}$.
}

\eqref{eqn: moment dynamics} determines the long-run behavior of the moments. For example, if $D = 0$, the covariance satisfies
\begin{equation} \label{eqn: covariance no mut}
    \Sigma(t) = (\Sigma(0)^{-1} - H_{xx} t)^{-1}.
\end{equation}
Thus, if $H_{xx}$ is negative semi-definite, $\Sigma(t) \to 0$ as $t \rightarrow \infty$. If $H_{xx}$ has any positive eigenvalues, then the covariance diverges, corroborating the intuition that the population should only concentrate at maxima of the local rating function $r$.

\textbf{Lemma 4: [Covariance Dynamics Sans Mutation] } \textit{If $f$ is quadratic, $\pi$ is multivariate normal and governed by the continuous replicator equation without mutation, then when $H_{xx}$ is negative semi-definite the covariance $\Sigma(t)$ is defined for all time, and converges to zero at rate $\mathcal{O}(t^{-1})$. Otherwise, the covariance diverges in finite time, at the first $t$ such that an eigenvalue of $I - \Sigma(0) H_{xx} t$ crosses zero.} 

Therefore, without mutation to support a diverse population, the population collapses to a monomorphism at $\bar{x}_*$ where the optimal self-play training field, $J \bar{x}_* + g$ equals $0$. This is a clear example of a concentration mechanism. In fact, direct perturbation analysis yields:
\begin{equation}
    \lim_{t \rightarrow \infty} t \Sigma(t) = -H_{xx}^{-1} + \mathcal{O}(t^{-2})
\end{equation}
and, for large $\tau$:
\begin{equation} \label{eqn: approximate centroid solution}
    \bar{x}(\tau + t) \simeq \bar{x}_* + t^{-I - H_{xx}^{-1} H_{xy}} (\bar{x}(\tau) - \bar{x}_*).
\end{equation}
 If $H_{xx}$ is negative definite, all the eigenvalues of $-I - H_{xx}^{-1} H_{xy}$ have negative real parts, so $\bar{x}(\tau + t)$ converges to $\bar{x}_*$. Note that, at $\bar{x}_*$, the self-play gradient vanishes, so the local linearization is flat. In this case, the population concentrates over time, but the self-play gradient also vanishes over time. This example demonstrates the need for the careful asymptotic theory developed in Section \ref{sec: convergence} and \cite{cebra2023similarity}. If the covariance vanishes fast enough, then the vanishing linear model will dominate, producing hierarchical systems. If, instead, the covariance vanishes slow enough, then cyclicity may survive.


In contrast, consider the case with mutation. The diffusive term $D$ constantly introduces diversity, so the covariance reaches a non-zero steady state at $\Sigma_*$ satisfying:
\begin{equation} \label{eqn: covariance steady state}
    \Sigma_*^{\intercal} H_{xx} \Sigma_* = - D.
\end{equation}

\eqref{eqn: covariance steady state} can be solved using the singular value decompositions of $D$ and $H_{xx}$. In particular:
\begin{equation} \label{eqn: steady state bound}
    \sqrt{\frac{\sigma_1(D)}{\sigma_{1}(H_{xx})}} \leq \|\Sigma_*\|_2 \leq \sqrt{\frac{\sigma_1(D)}{\sigma_{\textnormal{min}}(H_{xx})}} 
\end{equation}
where $\sigma_j(A)$ denotes the $j^{th}$ singular value of $A$. Since $H_{xx}$ is negative definite, \eqref{eqn: steady state bound} ensures that, in the long run, $\pi(t)$ neither goes to a delta nor diverges. Nevertheless, concentration results provide useful approximation when $\Sigma_*$ is small, and, exact predictions when $f$ is quadratic.

To compare the degree of cyclicity retained with and without mutation, compute $\epsilon$. For $D = 0$, the difference between the rates of concentration and convergence of $\bar{x}(t)$ to $\bar{x}_*$ determines the long-term behavior of the system. If $H_{xx}$ is negative semi-definite, $\epsilon$ remains nonzero: 
\begin{equation} \label{eqn: epsilon no mutation}
\begin{aligned}
\lim_{t \rightarrow \infty} \epsilon(t) 
=  \frac{\langle H_{xy} H_{xx}^{-1},H_{xx}^{-1} H_{xy} \rangle}{\langle H_{xx} H_{xx}^{-1},H_{xx}^{-1} H_{xx} \rangle} 
& = \frac{\langle H_{xy} H_{xx}^{-1},H_{xx}^{-1} H_{xy} \rangle}{T}
\end{aligned}
\end{equation}
Therefore, some intransitivity survives the concentration limit. 

If $D \neq 0$, then the long-term covariance approaches a steady state $\Sigma_*$ and the long-term mean converges to $\bar{x}_*$ where the local linear model vanishes. For large $t$, we can approximate $\Sigma(t)$ by $\Sigma_*$. Therefore,
\begin{equation} \label{eqn: epsilon mutation}
    \lim_{t \rightarrow \infty} \epsilon(t) = \frac{\langle H_{xy} \Sigma_* , \Sigma_* H_{xy} \rangle}{\langle H_{xx} \Sigma_* , \Sigma_* H_{xx} \rangle}
\end{equation}

Thus, under mutation, the steady-state intransitivity is independent of the magnitude of $D$. Note that, intransitivity survives the long time limit with and without mutation, however, the ratios determining the ultimate degree of cyclicity differ based on the limiting population covariance. 

\subsection{Comparative Steady-State Cyclicity Analysis} \label{sec: steady state cyclicity}

Equations (\ref{eqn: epsilon no mutation}) and (\ref{eqn: epsilon mutation}) provide two examples of the steady state $\epsilon$ under different limiting scenarios and learning rules. More generally, when $f$ is quadratic, $\pi$ is Gaussian, and the self-play gradient vanishes at the centroid, then:
\begin{equation} \label{eqn: Gaussian epsilon}
    \epsilon = \frac{\langle H_{xy} \Sigma, \Sigma H_{xy} \rangle }{\langle H_{xx} \Sigma, \Sigma H_{xx} \rangle}  = \frac{\|\tilde{H}_{xy}\|_{\text{Fro}}^2}{\|\tilde{H}_{xx}\|_{\text{Fro}}^2}
\end{equation}
where $\tilde{H}$ denotes the Hessian after a change of coordinates that whitens the steady state covariance $\Sigma$ \cite{cebra2023similarity}. 

\eqref{eqn: Gaussian epsilon} is useful for comparing the steady state intransitivity of different learning rules. Different learning rules induce different steady state covariances. For example:
\begin{enumerate}
    \item \textbf{Logit: } Under the logit dynamic, agents switch strategies according to a softmax distribution applied to the expected payout of each strategy \cite{lahkar2015logit}. This induces a logit equilibrium; the steady state log density must be proportional to the expected payout at each trait \cite{anderson2002logit,anderson2004noisy}. For quadratic $f$, the steady state is Gaussian with covariance $\Sigma_* \propto -H_{xx}^{-1}$ as posited in \cite{cebra2023similarity}. This is also the limit achieved by the continuous replicator without mutation.

    \item \textbf{Continuous Replicator + Diffusion: } $\Sigma_*$ solves \ref{eqn: covariance steady state}.

    \item \textbf{Noisy Self-Play (OU): } Consider a population consisting of many independent individuals, each learning through self-play with Gaussian noise arising either from errors in gradient evaluation, finite sampling, or as intentional exploration. Given quadratic $f$, the self-play gradient is a linear function of traits, so for fixed noise, the agents obey an Ornstein-Uhlenbeck process, and $\Sigma_*$ must satisfy the Lyapunov equation: $\Sigma_* J + J \Sigma_* = - D$ where $D$ is the noise covariance in each step \cite{van1992stochastic}.
\end{enumerate}

The three cases, $\Sigma_* \propto -H_{xx}^{-1}$, $\Sigma_* H_{xx} \Sigma_* = - D$, and $\Sigma_* J + J \Sigma_* = - D$ produce different $\epsilon$ thus degrees of cyclicity. 

The different steady-state equations can be used to compare the degree of cyclicity supported by different learning rules, random matrix models, or random models for $f$. For example, when $f$ is drawn from a Gaussian process with a smooth, stationary kernel, then the Hessian at $\bar{x}_*$ has Gaussian entries \cite{mchutchon2015nonlinear,dean2008extreme,yamada2018hessian}. When the entries of the Hessian are Gaussian distributed, or when the Hessian blocks are drawn from unitarily invariant families, the numerators and denominators of $\epsilon$ can be computed in expectation using tools from random matrix theory \cite{collins2022weingarten,novak2014three}. 
More generally, given a random matrix family, one can easily compare steady-state cyclicity experimentally. 

\begin{figure*}[h]
    \centering
    \includegraphics[scale = 0.35]{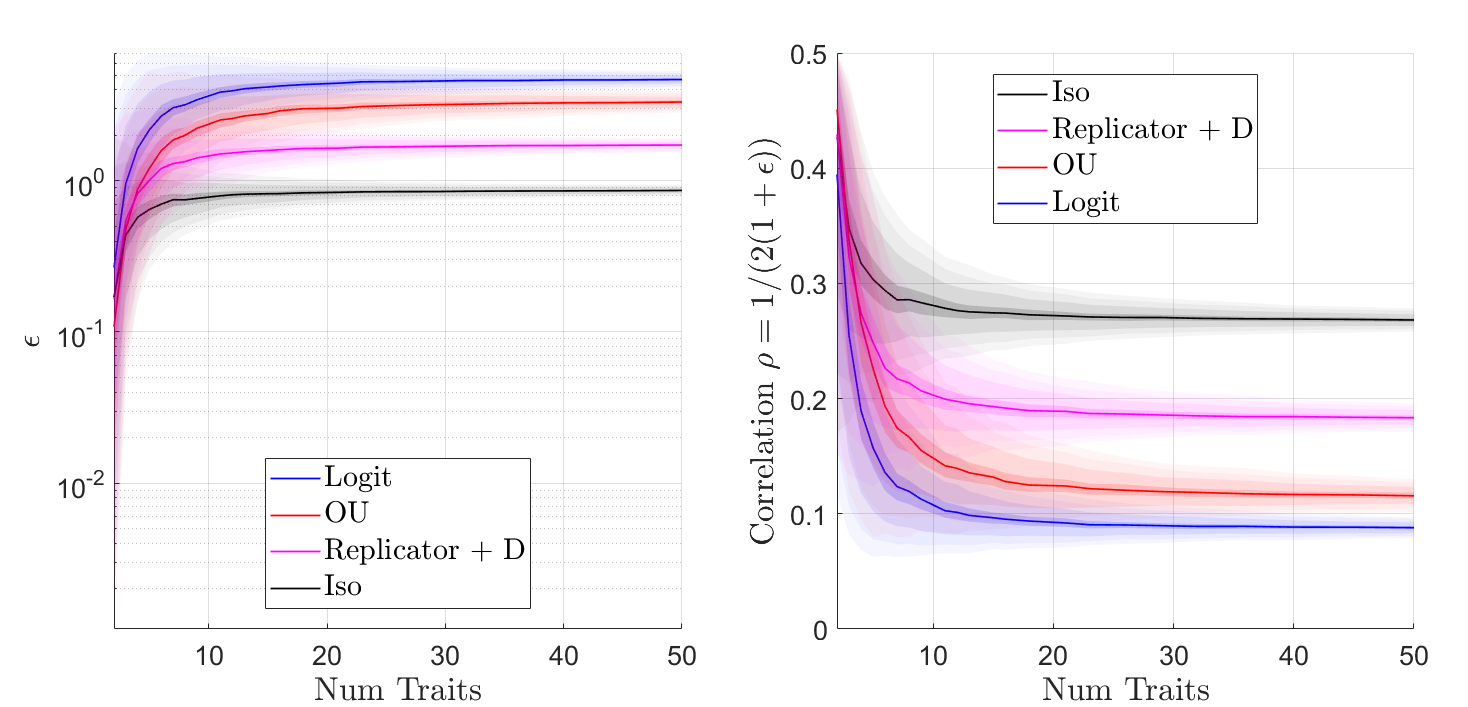}
    \caption{The ratio of moments $\epsilon$ and correlation $\rho$ for three learning rules, with Hessian blocks drawn from symmetric and skew-symmetric Gaussian families, and diffusion drawn as the product of Gaussian matrices. The isovariant case uses $\Sigma_x \propto I$ to establish a reference. We require that the $H_{xx}$ block is negative definite with largest eigenvalue $\lambda \leq -\sqrt{T}/10$. Sampling from the conditional distribution is performed using a mixed Langevin-Gibbs sampler \cite{yamada2018hessian}. The scaling ensures that the conditioning of $H_{xx}$ converges for large $T$, thus the eccentricity of the steady states converge. Median results are shown with solid lines while shaded regions represent the [5\%,95\%], [10\%,90\%],[20\%,80\%], and [40\%,60\%] intervals for 1,000 samples. Recall that $\rho = 0.5$ implies perfect transitivity}
    \label{fig: random matrix comparison}
\end{figure*}

 Figure \ref{fig: random matrix comparison} provides an example. In all tests, $\rho$ concentrated for large $T$, decreased with increasing steady state eccentricity, and the logit rule produced the most cyclic networks, while the isovariant reference produced the most transitive.  

\section{Numerical Demonstration} \label{sec: numerics}

To test our theory, we simulated the continuous replicator equation and a Gaussian adaptive process on a series of bimatrix games and random performance functions. We simulated the replicator equation to confirm our dynamical analysis and simulated the Gaussian adaptive process to test the generality of our results under other selection dynamics. In each case, we tested whether evolution promotes concentration towards a small subset of the strategy space, and, consequently, promotes transitivity at the convergence rates predicted by Theorem 1. 

\subsection{Example Games} \label{sec: numerics background}
Bimatrix games provide a simple example with well-documented Nash equilibria. For a review and taxonomy see \cite{bruns2015names,liebrand1983classification}. We tested three examples: chicken, prisoner's dilemma (PD), and stag hunt (Stag). Table \ref{tab: bimatrix games list} shows the associated payout matrices. 

PD presents a trust dilemma via a choice between cooperation and defection \cite{Skyrms}. Once iterated, it can model the evolution of altruism \cite{axelrod1981evolution}. Like PD, Stag is a coordination game \cite{Skyrms}. Each individual can independently hunt a hare for a small guaranteed payout or can choose to hunt a stag for a higher payout. It takes both players to catch the stag. The game of chicken models escalation and brinksmanship \cite{Rapoport}. Competitors choose to swerve or stay the course. If only one competitor stays, then they gain a small reward. If both stay, then they crash and die. 
\begin{table}[h!]
\begin{centering}
\begin{tabular}{ |p{1.8cm}| p{2cm}|p{4cm}|  }
 \hline
 \multicolumn{3}{|c|}{List of bimatrix games} \\
 \hline
 Games& Dilemma &Payout Matrix\\
 \hline
 Prisoner's dilemma &  Trust   & \begin{centering} $\begin{bmatrix}
(2,2) & (0,3)\\
(3,0) & (1,1)
\end{bmatrix}$ \end{centering}\\
 Stag hunt & Cooperation & $\begin{bmatrix}
(6,6) & (1,3)\\
(3,1) & (2,2)
\end{bmatrix}$\vspace{0.05in} \\
 Chicken   &  Escalation   & $\begin{bmatrix}
(1000,1000) & (999,1001)\\
(1001,999) & (0,0)
\end{bmatrix}$ \vspace{0.02in}\\
 \hline
\end{tabular}

 \end{centering}
 \vspace{2mm}
 \caption{List of bimatrix games considered and their payout matrices.}
\label{tab: bimatrix games list}
\end{table}

For each bimatrix game, the trait $x \in \Omega = [0,1]$ represents the probability an agent chooses the first action. To produce a skew-symmetric performance function that encodes both cooperative and competitive advantage, we replace individual agents with populations.  Then, we initialized a Moran process with equal proportions playing $x$ and $y$ and terminated at fixation to a single type \cite{lieberman}. We let $f(x,y)$ represent the probability $x$ fixes when competing against a population of type $y$, minus one half. To speed evaluation, we interpolated fixation probabilities simulated on an evenly spaced grid. 

To test the generality of our analysis in higher dimensional trait spaces where interactions between distinct traits of distinct agents can produce intransitivity \cite{cebra2023similarity}, we also used an ensemble of randomly generated performance functions. Specifically, we set $f$ to a sparse linear combination of Fourier and polynomial basis functions with random coefficients. By varying the coefficient distributions, we tuned the expected structure of $f$. Low-order terms produce nearly transitive functions on small neighborhoods, while high-order terms remain intransitive except at very small scales. For this example, we interpret $f$ as the log-odds of victory in a game with binary outcomes. For a full description, see the Supplement.

To test the replicator dynamic, we used a randomly chosen quadratic function $f$ in a 2-dimensional trait space. The function was chosen so that it admits an ESS at the origin, $H_{xx}$ was symmetric negative definite, and $H_{xy}$ was skew-symmetric.

\subsection{Population Dynamics} \label{sec: numerics procedure}

To test the generality of our results under alternative dynamics, we implemented multiple stochastic, discrete-time, agent-based evolutionary models. To initialize each process, we sampled a set of competitors uniformly from the trait space $\Omega$. The trait space was 1-dimensional for bimatrix games and $n$-dimensional for random performance functions. Then, at each epoch, every competitor played a series of games against a random selection of other competitors drawn without replacement from the current population. Game results were determined by the performance function. Next, we selected agents to reproduce based on a selection criteria. We tested logit and softmax functions of agents' average performance, and a cutoff that left only the top 10\% of competitors sorted by performance to reproduce. All individuals are removed at each generation and replaced by their children. Children inherit their parent's traits with a random Gaussian perturbation. We repeated the process until the centroid of the trait distribution of agents stopped moving appreciably. 

To measure the degree of transitivity at each epoch, we use the relative sizes of the perfectly transitive and cyclic components of the complete graph of competitors \cite{jiang2011statistical,strang2022network}. 

We perturbed the system in various ways--including varying the number of opponents each competitor played per epoch, the total number of competitors, and the degree of genetic drift per epoch. For the random performance functions, we also varied the number of traits, and the distribution used to sample the coefficients. For control parameters and a list of parameters tested, see the Supplement.

Finite populations whose agents switch strategies with differential rates determined by differences in average performance obey, in the limit of large population sizes, a mean-field replicator dynamic \cite{schuster}. To test the theory developed in section \ref{sec: replicator dynamic}, we simulated a large population of agents switching strategies through imitation learning. The starting competitors were sampled from an $n$-dimensional Gaussian trait distribution. Then, at each step, two randomly chosen competitors $i$ and $j$ compare their average performance against the full population, $f_i$ and $f_j$. Competitor $i$ switches to the strategy of competitor $j$ with probability $\frac{1}{2} + \frac{f_j-f_i}{2}$, and $j$ switches to $i$ according to the same rule with $i$ and $j$ reversed. Then, all competitors' trait vectors are perturbed by Gaussian noise.

On average, one switching event happens per evolutionary step. An epoch is the number of steps needed until the average number of switches matches the population size. To monitor the population distribution we track its centroid, covariance, skewness, and kurtosis and compare to predictions using the replicator equation. To measure intransitivity, we perform an HHD at each step, and compare to analytical estimates based on Gaussian approximation. 

\subsection{Replicator Results} \label{sec: replicator dynamic}


To test the replicator analysis, we initialized $2 \times 10^4$ agents away from the ESS with high genetic drift, then monitored the relaxation to steady state. The level of genetic drift was chosen so that the rate of noise-driven drift per epoch matched the rate of motion due to selection. This avoids an initial clustering phase during which the population collapses into competing clusters and allows comparison to the Gaussian limit. To test Gaussianity, we monitored the skewness and kurtosis which remained close to 0 and 3 respectively. Thus, the family of approximately Gaussian distributions is stable under the replicator dynamic with diffusive noise and quadratic performance. Even when using a finite population, Gaussian-distributed populations remain Gaussian.

\begin{figure}[h]
    \centering
    \includegraphics[scale=0.45]{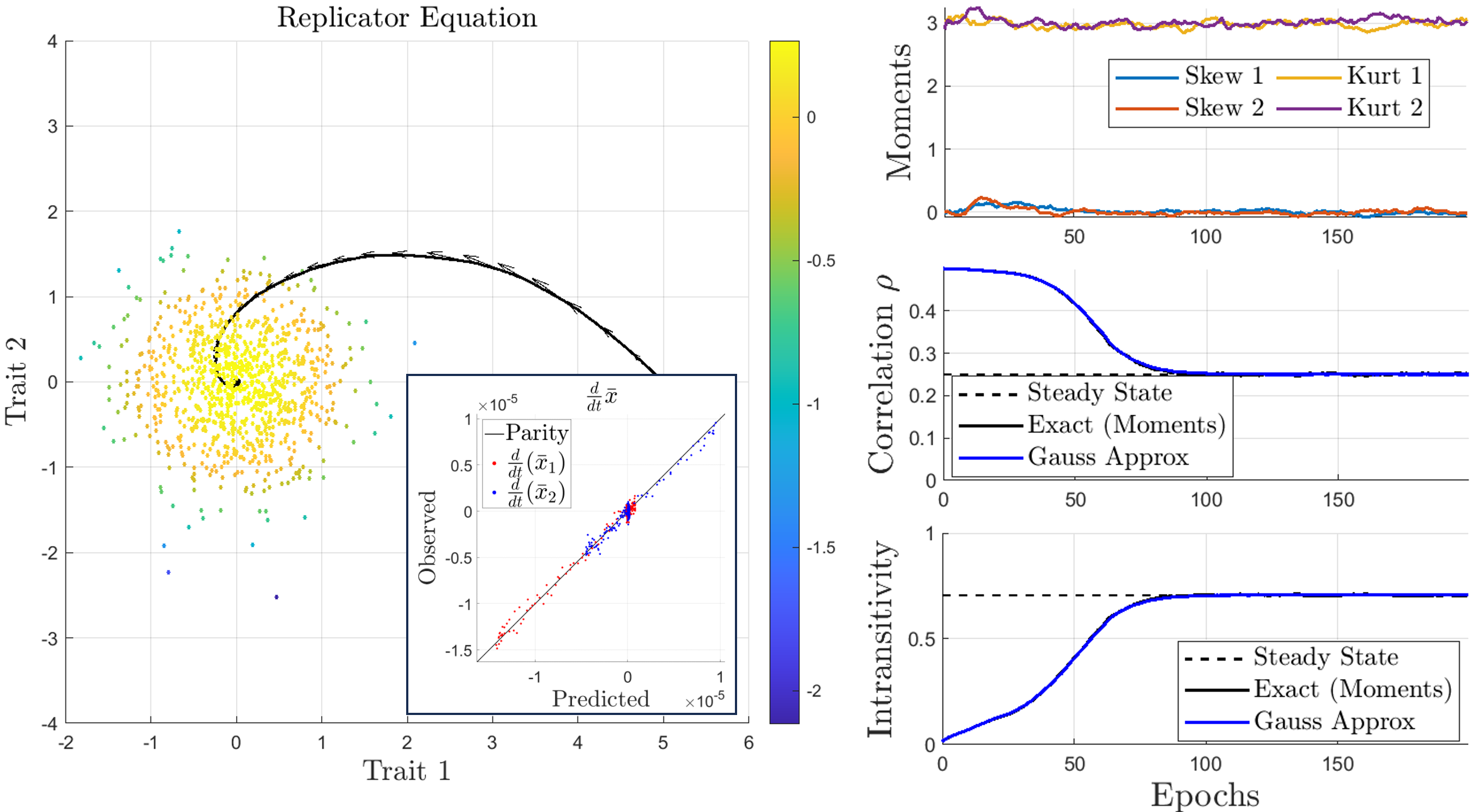}
    \caption{Results from the replicator example. \textbf{Left: } The trajectory of the centroid with tangent arrows for a performance function with a peak at the origin, where all competitors are selected to start centered at a point away from the origin. \textbf{Left inset: } For each epoch, predicted movement of the centroid according to the replicator equation compared to actual movement. \textbf{Top right: } Moments of the trait distribution per epoch, per dimension. \textbf{Middle right: } Covariance of the trait distribution per epoch. \textbf{Bottom right: } Relative intransitivity, per epoch. For a video animation of the trait distribution over time, see \url{https://github.com/ccebra/Evolution-Almost-Sure-Hierarchy/tree/main}}
    \label{fig:replicatortrajectory}
\end{figure}

To check the moment dynamics, we compared the predicted motion of the centroid under a Gaussian approximation, to the realized motion of the centroid. Figure \ref{fig:replicatortrajectory} compares the realized trajectory to the centroid dynamics (grey arrows). Note that, at all stages, the predicted direction of motion is nearly tangent to the realized motion, verifying equation \eqref{eqn: moment dynamics}. Similar tests support the covariance dynamics but require a larger population for close agreement. Nevertheless, the realized covariance converged to the steady state predicted using \eqref{eqn: covariance steady state} (the dashed line).

Finally, we compared the realized correlation coefficient $\rho$ and intransitivity to their Gaussian population approximation. At all epochs, the correlation and intransitivity closely matched their Gaussian approximation \cite{cebra2023similarity} (see Figure \ref{fig:replicatortrajectory}). Moreover, the realized correlation and intransitivity converged to their predicted steady-state values using \eqref{eqn: epsilon mutation} (dashed line).

\subsection{Bimatrix Results} \label{sec: bimatrix}

\begin{figure*}[h]
		\centering
		\includegraphics[scale=0.55]{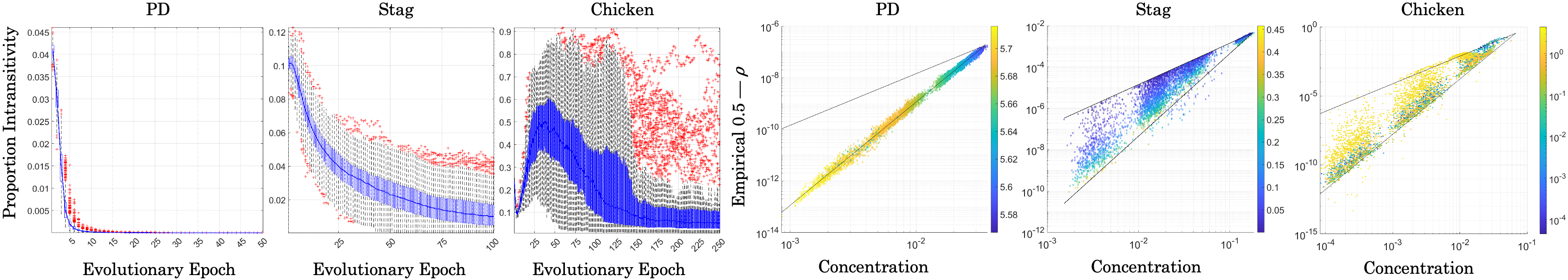}
		\caption{\textbf{Left: } Step-by-step intransitivity for the three bimatrix games under control parameters. \textbf{Right: } The empirical correlation $\rho$ versus concentration $\kappa$ (trait-averaged standard deviation) in the trait distribution. Colors represent the norm of the gradient of the performance function, $g$, for PD and Stag, and $H_{xx}^2/(g  \text{ Cov})^2$ for chicken, all evaluated at the final centroid. The lines mark quartic ($\mathcal{O}(\kappa^4)$) and quadratic convergence ($\mathcal{O}(\kappa^2)$) from the maximum empirical $\rho$. Data is generated using repeated trials with varying genetic drift to produce clusters with varying concentration. Trials with multiple clusters at the end of evolution (all in chicken) have been removed.}
		\label{fig: step by step bimatrix}
\end{figure*}

Figure \ref{fig: step by step bimatrix} shows the intransitivity per epoch for each bimatrix game under control parameters. The transitivity dynamics varied by example. In PD, intransitivity dropped rapidly to zero. In Stag, intransitivity converged to approximately 1\%. In Chicken, intransitivity grew rapidly, before decaying to roughly 10\%. Thus, chicken sustained appreciable intransitivity. The speed with which each experiment converges is also notable. In PD, the population reached the Nash equilibrium (NE) in fewer than 10 epochs, whereas in Stag the process took around 100 and in Chicken it takes nearly 250. 
Stag and Chicken both support an interior NE while PD only supports a NE on the boundary of the strategy space.
The final strategy distributions and NE are available in the Supplement.

Most of the parameters varied had little to no effect on the transitivity dynamics, and, in all cases, intransitivity decreases with time. In contrast, the degree of genetic drift --- the standard deviation between the traits of a parent agent and its children --- was impactful.  The genetic drift determines how much the population explores and how tightly it concentrates. 
For PD, intransitivity is insensitive to the genetic drift parameter since selection to the boundary is very strong. For Stag, increasing drift increases the final proportion of intransitivity by increasing the diversity in the final population. Results using Chicken are highly susceptible to the genetic drift. When drift is 10 times larger than control, the degree of intransitivity remains around 0.4 rather than returning to 0.1.

Figure \ref{fig: step by step bimatrix} illustrates the rate of convergence to transitivity as a function of concentration at the final epoch. To generate examples with varied degrees of concentration, we varied the drift parameter. 
We compared these results to the predictions in Table \ref{tab: rate of convergence}. In all cases, empirical results matched prediction. In PD, the NE lay on the boundary and is supported by a nonvanishing gradient selecting for agents near the boundary. Thus, convergence is quartic. In Stag, some trials ended with the centroid removed from the interior NE. Then, the gradient is nonzero producing quartic convergence. If the centroid finds the NE, the gradient is 0 producing quadratic convergence. For Chicken, higher-order derivatives dominated since $f(x,y)$ has a sharp cusp near $x,y = 0,0$. As a result, rates depended on the ultimate location of the centroid; some trials converged quartically (see the blue scatter points in Figure \ref{fig: step by step bimatrix}). Others decayed quadratically (see the yellow band).

\subsection{Random Performance Results} \label{sec: random performance functions}

Next, we repeated our experiment using random $f$ in multi-dimensional trait spaces.

\begin{figure*}[h] 
		\centering
		\includegraphics[scale=0.51]{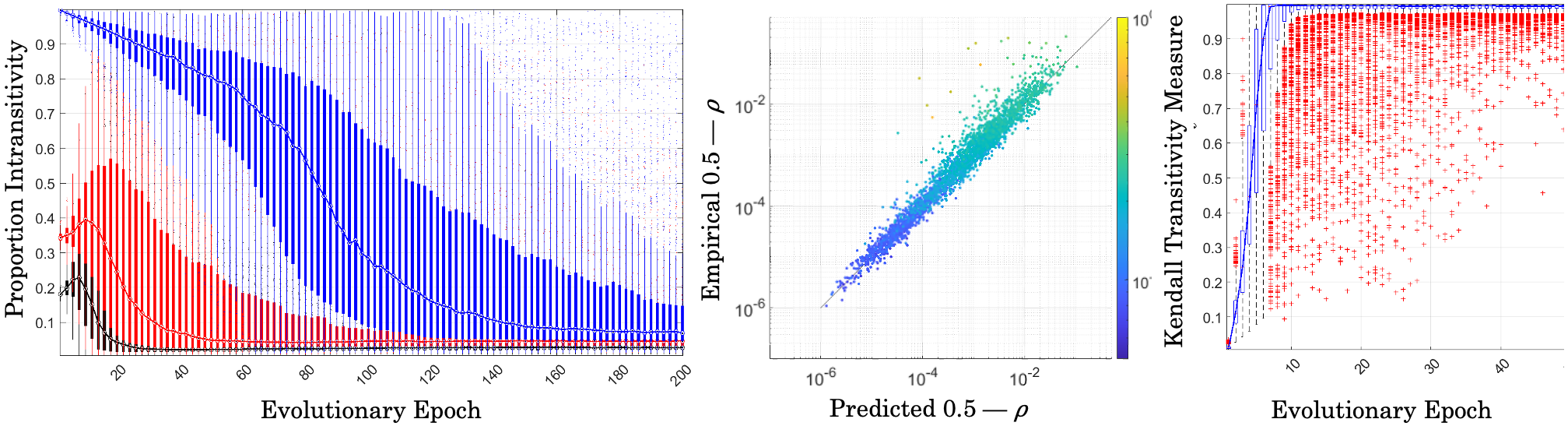}
		\caption{Trait concentration and transitivity for random performance functions in 4-dimensional trait spaces under control parameters under the softmax selection criterion. \textbf{Left: } Intransitivity by step according to the HHD for random performance functions with different trigonometric and linear weights. (Black): Equal weighting of trigonometric and linear. (Red): Trigonometric has double weight over linear. (Blue): Function is entirely trigonometric. \textbf{Middle: } Correlation versus predicted. \textbf{Right: } Kendall transitivity measure by step for control parameters.} 
		\label{fig: control parameters step by step}
\end{figure*}

Figure \ref{fig: control parameters step by step} summarizes the results. Under control parameters intransitivity quickly vanished, since the covariance in the population converged to zero rapidly.  
%
The middle panel shows the close match between observed intransitivity and predictions based on second-order Taylor expansion of $f$ about the realized centroid, assuming a Gaussian distribution (see Table \ref{tab: rate of convergence} and \cite{cebra2023similarity}). As predicted, the observed intransitivity vanished quadratically with concentration. 
The right panel shows that the Kendall transitivity measure reached 1 for most experiments, supporting the conclusion that, if an ensemble converges to perfect transitivity in expectation, then it converges to transitivity in probability.

To test how transitivity dynamics depend on the roughness of $f$, we altered the relative amplitudes of its linear and trigonometric components. The trigonometric components are responsible for all nonlinearities, and thus intransitivity, in our performance function. Decreasing their relative amplitude should promote transitivity. The left panel of figure \ref{fig: control parameters step by step} illustrates this effect. Even when starting from a purely trigonometric $f$, which created a perfectly \textit{intransitive} network before evolution, concentration during evolution suppressed intransitivity near to zero (6\% at the last epoch).

We also tested a variety of different selection mechanisms as described in Section \ref{sec: numerics procedure}. In general, the top 10\% cutoff model produced the fastest convergence towards transitivity, due to rapid concentration. The softmax and logit models exhibited more gradual convergence towards transitivity.

While changing the parameters or model changed how quickly the populations approached perfect transitivity, the only perturbation that prevented convergence to transitivity was increasing genetic drift, as, for large genetic drift, populations cannot concentrate sufficiently (see supplement). 

\section{Discussion} \label{sec: discussion} 

The mechanism developed here is limited by its assumptions. It cannot explain hierarchy among diverse populations, or for rough payout functions. Moreover, it only applies if the set of relevant agent pairs is independent of agent attributes, and thus, performance \cite{strang2022network}. The latter assumption is strongly violated in certain communities \cite{chase,Chase_b,drummond,dugatkin,hogeweg,Hsu,kawakatsu,Oliveira,Silk,vanhonk} but serves as a useful null hypothesis for comparison. Further social effects likely compound the concentration effect. 

Future work could build on the theory developed here. In particular, the random matrix study presented in Section \ref{sec: steady state cyclicity} merits a systematic investigation, validation, and comparison with varying random models for the constituent matrices. Further work could compare the expected degree of hierarchy by studying $\rho$ for different random functions. In particular, the sampling distribution for $\rho$ under Gaussian process models could be used to develop novel hypothesis tests for transitivity, and to better understand the interaction between smoothness, diversity, and cyclicity beyond local approximation. 

We are grateful to Karen Abbott, Dane Taylor, Hye Jin Park, and Yuriy Pichugin for discussing this work.


\vspace{6.7 in}

\pagebreak
\section*{Supplementary Analysis}
\subsection{Proof of Theorems 2 and 3 in Section C}
Here, we give a full proof of the arguments that lead to the Almost Sure Transitivity result (Theorem 3) in Section C of the main text.

First we aim to reduce the study of generic networks to the study of triangles, then to a statement regarding $f_{ij}$ on a random $i,j$ pair. This argument avoids topological issues and focus on a simple probabilistic statement on an individual edge. Namely, if $X,Y$ are the (randomly sampled) traits of two competitors drawn i.i.d.~from $\pi_x(\kappa)$ then, does the probability $|f(X,Y)| \leq \alpha(\kappa)$ vanishes for a bound $\alpha(\kappa)$ that converges to zero slower than $\kappa^4$? We establish a smoothness condition on $f$, and concentration condition on $\pi_x(\kappa)$ for which $|f(X,Y)|$ almost always vanishes slowly enough, thus proving that, for any $z$ such that $\nabla_x f(x,y)|_{x=y=z} \neq 0$, there exists a concentration parameter $\kappa$ small enough such that a competitive network sampled from $\pi_x(\kappa)$ is almost always transitive. If, as is generically true, $f(x,y)$ is nonconstant on open sets, then, for almost all $z \in \Omega$, $\nabla_x f(x,y)|_{x=y=z} \neq 0$, so, for almost all $z \in \Omega$, there is a distribution, sufficiently concentrated about $z$, such that competitive systems sampled from the distribution are transitive with probability arbitrarily close to one.

\subsubsection{Reduction to Argument on a Single Edge}

Consider a trait space $\Omega$, equipped with a trait distribution, $\pi_x(\kappa)$, which depends on a concentration parameter $\kappa$. Then, consider an ensemble of $V$ agents with traits $X_i \sim \pi_x(\kappa)$ drawn i.i.d. We let $X$ denote the set of trait vectors. Then, introducing a performance function $f$, and a set of competitors who could compete, $\mathcal{E}$, produces a competitive network $\mathcal{G}(X,f,\mathcal{E})$. Let $\mathcal{K}(X,f)$ denote the associated complete graph, where all competitors can compete.

\vspace{2mm}
\noindent
\textbf{Lemma 1: } \textit{The probability $\mathcal{G}(X,f,\mathcal{E})$ is transitive is greater than or equal to the probability that $\mathcal{K}(X,f)$ is transitive.}

\vspace{2mm}
\noindent \textbf{Proof: } Transitivity is a condition on the set of cycles in a competitive network. The complete graph $\mathcal{K}(X,f)$ contains more cycles than the original network, $\mathcal{G}(X,f,\mathcal{E})$, and contains all of the cycles in $\mathcal{G}(X,f,\mathcal{E})$. Since moving from $\mathcal{G}(X,f,\mathcal{E})$ to $\mathcal{K}(X,f)$ does not change the advantage on any edge, $\mathcal{K}(X,f)$ is only transitive if $\mathcal{G}(X,f,\mathcal{E})$ is transitive (no loops in $\mathcal{G}_{\rightarrow}(X,f,\mathcal{E})$), and there are no advantage loops amongst the additional cycles introduced by adding edges. Thus, adding edges can make a transitive network intransitive, but cannot make an intransitive network transitive. Therefore:
\begin{equation}
    \text{Pr}\{\mathcal{G}(X,f,\mathcal{E}) \text{ is transitive} \} \geq \text{Pr}\{\mathcal{K}(X,f) \text{ is transitive} \} \quad 
\end{equation}

In other words, the set of graphs whose completion is transitive is contained in the set of transitive graphs. $\square$

\vspace{2mm}

Lemma 1 allows us to ignore $\mathcal{E}$, and focus only on the traits of the sampled competitors, $X$, and performance function $f$. We use the probability that the complete network is transitive as a lower bound on the probability the original network is transitive. Working with complete graphs is convenient, since a complete graph is transitive if and only if all triangles in the graph are transitive. Thus, we can reduce our arguments from arbitrary networks, to arbitrary complete networks, to arbitrary triangles. 

\vspace{2mm}
\noindent \textbf{Lemma 2: } \textit{Let $\triangle_{i,j,k}(X,f)$ denote the triangular competitive network formed by distinct agents $i$, $j$, and $k$. Then, the probability that $\mathcal{K}(X,f)$ is transitive is greater than or equal to the probability that $\triangle_{i,j,k}(X,f)$ is transitive, for any $i \neq j \neq k$.}
\vspace{2mm}

\noindent \textbf{Proof: } The proof is, again, trivial. Fix $i \neq j \neq k$. Then:
\begin{equation}
    \text{Pr}\{\mathcal{K}(X,f) \text{ is transitive} \} = \text{Pr}\{\triangle_{i,j,k}(X,f) \text{ is transitive} \} \text{Pr}\{\mathcal{K}(X,f) \text{ is transitive} | \triangle_{i,j,k}(X,f) \text{ is transitive}  \}.
\end{equation}

The conditional probability that the complete graph is transitive given that one of its triangles is transitive is less than or equal to one, so:
\begin{equation}
    \text{Pr}\{\mathcal{K}(X,f) \text{ is transitive} \} \geq \text{Pr}\{\triangle_{i,j,k}(X,f) \text{ is transitive} \}. \quad \square
\end{equation}

\vspace{2mm}

Like Lemma 1, Lemma 2 simplifies the problem by reducing our focus from generic networks to triangles. Since all of the competitor traits are sampled i.i.d.~from $\pi_x(\kappa)$, and the advantage relations are a deterministic function $f$ of the traits, any triangle drawn from the complete graph is statistically identical to any other triangle. That is the marginal distributions of possible competitive networks on each triangle are all identical. Thus, we can drop the $i,j,k$ subscript, and consider an arbitrary triangular network consisting of only three sampled competitors drawn i.i.d.~from $\pi_x$. Let $\triangle(X,f)$ denote such a network.\footnote{Note: it is equivalent to consider $\triangle(X,f)$ a randomly sampled triangle from $\mathcal{K}(X,f)$. }

\vspace{2mm}
\noindent \textbf{Corollary 2.1: } \textit{If,} $\text{Pr}\{ \triangle(X,f) \text{is transitive}\} \rightarrow 1$ \textit{as $\kappa \rightarrow 0$, then} $\text{Pr}\{ \mathcal{G}(X,f,\mathcal{E}) \text{is transitive}\} \rightarrow 1$ \textit{as $\kappa \rightarrow 0$.}
\vspace{2mm}

\noindent \textbf{Proof: } The proof is entirely trivial, and follows by chaining the inequalities in Lemmas 1 and 2. $\square$

\vspace{2mm}

We have now reduced the problem for generic networks to a problem on triangles. Triangles are easy to work with, since transitivity is guaranteed by a simple bound on the norms of the components $F_c$ and $F_t$, where $F$ is the advantage edge flow, $F_{i,j} = f(X_i,X_j)$. Namely, $\triangle(X,f)$ is transitive if:
\begin{equation}
    \| F_c \|^2 \leq \frac{1}{2} \|F_t\|^2. 
\end{equation}

The proof is provided in the supplement to \cite{strang2022network}.

Recall that, since the perfectly transitive and perfectly cyclic subspaces are orthogonal complements:
\begin{equation}
    \|F_t\|^2 +\|F_c\|^2 = \|F\|^2.
\end{equation}

Then, it is a simple algebra exercise to show that:
\begin{equation}
    \|F_c\|^2 \leq \frac{1}{3} \|F\|^2 \text{ implies } \| F_c \|^2 \leq \frac{1}{2} \|F_t\|^2. 
\end{equation}

\vspace{2mm}
\noindent \textbf{Lemma 3} \textit{A triangular competitive network with advantage edge flow $F$ is transitive if $\|F_c\|^2 \leq \frac{1}{3} \|F\|^2$.}
\vspace{2mm}

A triangular competitive network may be transitive even if $\|F_c\|^2 \geq \frac{1}{3} \|F\|^2$. Therefore:

\vspace{2mm}
\noindent \textbf{Corollary 3.1: } Let $\triangle(X,f)$ denote a triangular competitive with randomly sampled competitors drawn i.i.d.~from $\pi_x$, and let $F$ denote the associated edge flow. Then, the probability a competitive network $\mathcal{G}(X,f,\mathcal{E})$ is transitive is greater than or equal to the probability that $\|F_c\|^2 \leq \frac{1}{3} \|F\|^2$.
\vspace{2mm}

Corollary 3.1 is useful since our concentration theorem establishes a characteristic rate at which $\mathbb{E}[\|F_c\|^2]$ converges to zero as $\kappa$ goes to zero. Moreover, if $\nabla_x f(x,y)|_{x=y=z} \neq 0$, then $\mathbb{E}[\|F_c\|^2]$ goes to zero faster than $\mathbb{E}[\|F\|^2]$, so it is reasonable to expect that, with high probability, $\|F_c\|^2 \leq \frac{1}{3} \|F\|^2$. 

To use our concentration theorem, we introduce a bound $\alpha(\kappa)$ designed to separate $\|F_c\|^2$ and $\|F\|^2$. Namely, $\alpha(\kappa)$ is meant to act as an upper bound on $\|F_c\|^2$, and a lower bound on $\frac{1}{3}\|F\|^2$. If $\alpha(\kappa)$ converges to zero slower than $\mathbb{E}[\|F_c\|^2]$, but faster than $\mathbb{E}[\|F\|^2]$, then it is reasonable to hope that $\|F_c\|^2 \leq \alpha(\kappa) \leq \frac{1}{3} \|F\|^2$, ensuring transitivity.

\vspace{2mm}
\noindent \textbf{Lemma 4: } \textit{If there exists a real-valued function $\alpha(\kappa) \geq 0$ which converges to zero as $\kappa$ goes to zero such that:}
\begin{equation}
    \text{Pr}\{\|F_c\|^2 < \alpha(\kappa) \} \rightarrow 1 \text{ and } \text{Pr}\{\|F\|^2 > 3 \alpha(\kappa) \} \rightarrow 1 
\end{equation}
\textit{as $\kappa$ goes to zero, then:}
\begin{equation}
    \text{Pr}\{\mathcal{G}(X,f,\mathcal{E}) \text{ is transitive}\} \rightarrow 1.
\end{equation}
\vspace{2mm}

\noindent \textbf{Proof: } Suppose that such an $\alpha(\kappa)$ exists. Then: 
\begin{equation}
    \begin{aligned}
    \text{Pr}\left\{\|F_c\|^2 < \frac{1}{3} \|F\|^2 \right\} & \geq \text{Pr}\left\{\|F_c\|^2 < \alpha(\kappa) \text{ and }  \alpha(\kappa) < \frac{1}{3}\|F\|^2 \right\} = 1 - \text{Pr}\left\{\|F_c\|^2 > \alpha(\kappa) \text{ or }  \alpha(\kappa) > \frac{1}{3}\|F\|^2\right\} \\ & \geq 1 - \left(\text{Pr}\left\{\|F_c\|^2 > \alpha(\kappa)\right\} + \text{Pr}\left\{\alpha(\kappa) > \frac{1}{3}\|F\|^2 \right\} \right).
    \end{aligned}
\end{equation}

By assumption the latter probabilities converge to zero as $\kappa$ goes to zero, so $\text{Pr}\{\|F_c\|^2 < \frac{1}{3} \|F\|^2 \} \rightarrow 1$, in which case, by Corollary 2.1, the probability that $\mathcal{G}(X,f,\mathcal{E})$ converges to one. $\square$
\vspace{2mm}

The first condition requires that the cyclic component of a randomly sampled triangular network is not too large. The second requires that the overall set of advantages is not too small. The former is easy to guarantee using the concentration theorem \cite{cebra2023similarity}. That theorem is repeated here.

\vspace{2mm}
\textbf{Theorem 1: [Trait Concentration]}
\textit{Suppose that $f(x,y)$ is a bounded performance function on $\Omega \times \Omega$, and satisfies the following smoothness assumption: $f(x,y)$ is second differentiable at $x = y = z$ and the second order Taylor expansion of $f(x,y)$ about $x = y = z$ has errors that, on some ball centered at $z$ with radius $r(z) > 0$, are bounded by a power series of $(x - z)$ and $(y-z)$ whose lowest order terms are cubic. Suppose that $\pi_x$ has centroid $z$ and depends on a concentration parameter $\kappa$ such that the probability $p(\kappa)$ of sampling $X$ outside a ball $B_{R(\kappa)}(z)$ goes to zero  at rate $\mathcal{O}_{<}(\kappa^4)$ for some $R(\kappa)$ which converges to zero at rate $\mathcal{O}_{<}(\kappa^{4/5})$.}

\textit{Then $\epsilon$ and $\mathbb{V}_{X,Y}[f(X,Y)]$ converge to their approximations using the local quadratic model of $f(x,y)$ about $x = y = z$ with errors vanishing faster than the approximations terms so:}
$$    
\begin{aligned}
    & \mathbb{E}[||F||^2] = \mathcal{O}_{=}(\kappa^2) \\
    & \mathbb{E}[||F_t||^2] = \mathcal{O}_{=}(\kappa^2)\\
    & \mathbb{E}[||F_c||^2] = \mathcal{O}_{\leq}(\kappa^4)
\end{aligned}
$$
\textit{if $\nabla_x f(x,y)|_{x=y=z} \neq 0$ and with equality if and only if $H_{xy}(z,z) \neq 0$. }

\textit{If $\nabla_x f(x,y)|_{x=y=z} = 0$  and $H_{xx}(z,z) \neq 0$ then:}
$$
\begin{aligned}
    & \mathbb{E}[||F||^2] = \mathcal{O}_{=}(\kappa^4) \\
    & \mathbb{E}[||F_t||^2] = \mathcal{O}_{=}(\kappa^4) \\
    & \mathbb{E}[||F_c||^2] = \mathcal{O}_{\leq}(\kappa^4) \\
\end{aligned}
$$
\textit{with equality if and only if $H_{xy}(z,z) \neq 0$. }
\vspace{2mm}

See \cite{cebra2023similarity} for details. Thus, under the assumptions of the trait concentration theorem, $\mathbb{E}[\|F_c\|^2]$ vanishes at rate $\kappa^4$ or faster when $\nabla_x f(x,y)|_{x=y=z} \neq 0$. Then, applying the Markov inequality:

\vspace{2mm}
\noindent \textbf{Corollary 1.1: } \textit{Under the assumptions of Theorem 1, if $\nabla_x f(x,y)_{x=y=z} \neq 0$, then:}
\begin{equation}
    \text{Pr}\{\|F_c\|^2 \geq \alpha(\kappa) \} \rightarrow 0
\end{equation}
\textit{if $\alpha(\kappa) = \mathcal{O}_{>}(\kappa^4)$. }
\vspace{2mm}

\noindent \textbf{Proof: } The result is a simple application of Markov's inequality to the trait concentration theorem. By Markov's inequality:
\begin{equation}
    \text{Pr}\{\|F_c\|^2 \geq \alpha(\kappa) \} \leq \frac{\mathbb{E}[\|F_c\|^2]}{\alpha(\kappa)}.
\end{equation}

Under the assumptions of Theorem 1, if $\nabla_x f(x,y)|_{x=y=z} \neq 0$, then the numerator is $\mathcal{O}_{\leq}(\kappa^4)$. Therefore, the ratio converges to zero provided $\alpha(\kappa)$ converges to zero slower than $\kappa^4$, i.e.~$\alpha(\kappa) = \mathcal{O}_{>}(\kappa^4)$. $\square$
\vspace{2mm}

Therefore, the first condition in Lemma 4 is easily satisfied for $\alpha(\kappa)$ decaying slower than $\kappa^4$. Thus, we can reduce our original problem, ensuring transitivity with high probability, to a simple probabilistic inequality on $\|F\|^2$, which can be converted into an inequality on an individual edge:

\vspace{2mm}
\noindent \textbf{Theorem 2: } \textit{Suppose that the smoothness and concentration assumptions of Theorem 1 hold, and $\nabla_x f(x,y)|_{x=y=z} \neq 0$. If, in addition, there exists an $\alpha(\kappa) = \mathcal{O}_{>}(\kappa^4)$ such that:}
\begin{equation}
    \text{Pr}\{f(Y,W)^2 > 3 \alpha(\kappa) \} \rightarrow 1 
\end{equation}
\textit{for, trait vectors $Y,W$ drawn i.i.d.~from $\pi_x(\kappa)$ then:}
\begin{equation}
    \text{Pr}\{\mathcal{G}(X,f,\mathcal{E}) \text{ is transitive}\} \rightarrow 1
\end{equation}
\textit{as $\kappa \rightarrow 0$.}
\vspace{2mm}

\noindent \textbf{Proof: } Under the assumptions of Theorem 1, if the gradient $\nabla_x f(x,y)|_{x=y=z} \neq 0$, then Corollary 1.1 applies, so the first condition in Lemma 4 is satisfied. Thus, almost sure transitivity ($\text{Pr}\{\mathcal{G}(X,f,\mathcal{E}) \text{ is transitive}\} \rightarrow 1$) is guaranteed if $\|F\|^2 \geq \frac{1}{3} \alpha(\kappa)$ with high probability. 

Note that:
\begin{equation}
    \|F\|^2 = f(X_i,X_j)^2 + f(X_j,X_k)^2 + f(X_k,X_i)^2 \geq f(X_i,X_j)^2.
\end{equation}

Therefore, if, on a single edge, $f(X_i,X_j)^2 \sim f(Y,W)^2$ is greater than $3 \alpha(\kappa)$ with high probability, then so is $\|F\|^2$, ensuring the second condition of Lemma 4. $\square$
\vspace{2mm}

Thus, under the standard smoothness assumptions, and concentration assumptions, transitivity is assured in the concentration limit provided $ \text{Pr}\{f(Y,W)^2 > 3 \alpha(\kappa) \} \rightarrow 1$. That is, provided the advantage on any randomly sampled edge is not excessively small with high probability. Thus, the problem reduces to finding conditions that ensure the advantage on a randomly sampled edge is not too small with high probability. In other words, we need a lower bound on the advantage on a randomly sampled edge that holds with high probability in the concentration limit. 

\subsubsection{Probabilistic Lower Bounds on Advantage under Concentration }

We seek conditions under which $f(Y,W)^2$ is greater than a vanishing lower bound, $3 \alpha(\kappa)$, with high probability in the concentration limit. Although we know that, under the assumptions of Theorem 1, $\mathbb{E}[f(Y,W)^2]$ vanishes at rate $\mathcal{O}_{\leq}(\kappa^2)$, which is slower than the fastest possible decay rate of $\alpha(\kappa)$, we cannot use Theorem 1 alone to establish the required lower bound since it only puts an upper bound on the convergence rate of the expectation of $f(Y,W)^2$. We need a lower bound that holds with high probability. Such a bound will require a stronger concentration assumption. A more specific definition of concentration that puts a lower, as well as an upper, bound, on the rate at which the trait distribution concentrates is required. A sufficient definition follows.

\textbf{Definition: } A sequence of distributions $\pi_x(\kappa)$ \textit{concentrates to $z$ at rate $\mathcal{O}_{=}(h(\kappa))$} if $h(\kappa) \rightarrow 0$ as $\kappa \rightarrow 0$, and:
\begin{enumerate}
    \item for any $R(\kappa) = \mathcal{O}_{>}(h(\kappa))$, $\text{Pr}\{|X - z| > R(\kappa)\} \rightarrow 0$ for $X \sim \pi_x(\kappa)$, and
    \item for any sequence of nested sets $S(\kappa) \subseteq \Omega \subseteq \mathbb{R}^T$, $S(\kappa') \subseteq S(\kappa)$ if $\kappa' \leq \kappa$ , whose volume $\text{Vol}(S(\kappa))$ is $\mathcal{O}_{<}(h(\kappa)^T)$, $\text{Pr}\{X \in S(\kappa)\} \rightarrow 0$ as $\kappa \rightarrow 0$. 
\end{enumerate}
\vspace{2mm}

The first condition ensures that the trait distribution concentrates fast enough, that is, at least at rate $h(\kappa)$. The second ensures that the distribution does not collapse too quickly, nor can it be expressed as a combination of distributions collapsing at different rates. The latter case is important to rule out since we need to ensure that rare events associated with sampling $X$ in a vanishing set do not occur with finite probability when that set vanishes faster than the slowest rate of concentration of the distribution - as it is that slowest rate that sets an upper bound on the fastest $\alpha(\kappa)$ can decay. The latter condition is a stronger version of the requirement that the covariance is $\mathcal{O}_{=}(\kappa^2)$ used before, since it controls the limiting probability of sampling from sets that vanish too quickly, rather than constraining a particular moment. 

Suppose that $\pi_x(\kappa)$ concentrates to $z$ at rate $\mathcal{O}_{=}(\kappa)$. Then, write:
\begin{equation}
\begin{aligned}
    & \text{Pr}\{f(Y,W)^2 < 3 \alpha(\kappa)\} \leq \\
    & \hspace{1 cm} \text{Pr}\{f(Y,W)^2 < \alpha(\kappa) \text{ and } |Y - z| < R(\kappa) \text{ and } |W - z| < R(\kappa)\} + \text{Pr}\{|Y - z| > R(\kappa)\} + \text{Pr}\{|W - z| > R(\kappa)\}.
    \end{aligned}
\end{equation}

By assumption, the latter pair of probabilities converge to zero as $\kappa \rightarrow 0$, so, if $\text{Pr}\{f(Y,W)^2 < \alpha(\kappa) \text{ and } |Y - z| < R(\kappa) \text{ and } |W - z| < R(\kappa)\}$ converges to zero, so does $\text{Pr}(f(Y,W)^2 < 3 \alpha(\kappa))$, in which case $3 \alpha(\kappa)$ is a lower bound for $f(Y,W)^2$ almost surely. 

Define the sequence of nested sets:
\begin{equation}
    S(\kappa) = \{x,y \in \Omega \times \Omega| f(x,y)^2 < 3 \alpha(\kappa)\} \cap B_{R(\kappa)}(z) \times B_{R(\kappa)}(z)
\end{equation}
where $B_{R(\kappa)}(z)$ is the ball radius $R(\kappa)$ centered at $z$. Then, our goal is to show that $\text{Pr}\{X \in S(\kappa)\}$ goes to zero as $\kappa$ goes to zero. This is accomplished by putting an upper bound on the volume of $S(\kappa)$, then showing that this volume vanishes faster than the distribution concentrates. 

To bound the volume of $S(\kappa)$ we invoke the smoothness assumption on $f$ used in Theorem 1. This is reasonable, since Theorem 1 is required to ensure that $\|F_c\|$ vanishes quickly enough in the first place. 

Then, for any $z$ there exists a ball of nonzero radius $r(z) > 0$ centered at $z$ such that the second order Taylor expansion of $f$ only differs from $f$ by an error that is bounded by a power series whose lowest order terms are cubic. It follows that the discrepancy between $f$ and its local linear model about $z$ can be bounded by a power series whose lowest order terms are quadratic. Let $g(z) = \nabla_x f(x,y)|_{x=y=z}$ denote the gradient vector field. Then, as long as $r(z)$ is chosen finite, there exists a scalar $M < \infty$ such that:
\begin{equation} \label{eqn: error bound}
    |f(x,y) - g(z)^{\intercal} (x - y)| < M (\|x - z\| + \|y - z\|)^2
\end{equation}
for any $x,y \in B_{r(z)}(z) \times B_{r(z)}(z)$. 

Now, since $R(\kappa)$ converges to zero as $\kappa$ goes to zero, there is a $\kappa$ small enough such that $S(\kappa) \in B_{r(z)}(z) \times B_{r(z)}(z)$, in which case, Equation \ref{eqn: error bound} applies. Then:
\begin{equation}
    f(x,y) \in g(z)^{\intercal} (x - y) +[- M, M ] (\|x - z\| + \|y - z\|)^2.
\end{equation}

It follows that:
\begin{equation}
    |f(x,y)| \geq |g(z)^{\intercal} (x - y) | - M (\|x - z\| + \|y - z\|)^2.
\end{equation}

Therefore, $f(x,y)^2 > 3 \alpha(\kappa)$ if:
\begin{equation}
    |g(z)^{\intercal} (x - y) | - M (\|x - z\| + \|y - z\|)^2 > \sqrt{3 \alpha(\kappa)}
\end{equation}
or:
\begin{equation}
    |g(z)^{\intercal} (x - y) | > \sqrt{3 \alpha(\kappa)} + M (\|x - z\| + \|y - z\|)^2. 
\end{equation}

Note that $\|x - z\| \leq R(\kappa)$, and $\|y - z\| \leq R(\kappa)$ when constrained to $B_{R(\kappa)}(z) \times B_{R(\kappa)}(z)$, so $M (\|x - z\| + \|y - z\|)^2 \leq 2 M R(\kappa)$. Therefore, $|g(z)^{\intercal} (x - y) | > \sqrt{3 \alpha(\kappa)} + M (\|x - z\| + \|y - z\|)^2$ if:
\begin{equation}
    |g(z)^{\intercal} (x - y) | > \sqrt{3 \alpha(\kappa)} + 2 M R(\kappa)^2.
\end{equation}

Therefore $S(\kappa) \in C(\kappa)$ where $C(\kappa)$ is the covering set:
\begin{equation}
    C(\kappa) = \{x,y | |g(z)^{\intercal} (x - y) | < \sqrt{3 \alpha(\kappa)} + 2 M R(\kappa)^2  \} \cap B_{R(\kappa)}(z) \times B_{R(\kappa)}(z) 
\end{equation}
We call this set $C(\kappa)$ since it covers $S(\kappa)$, and its volume is bounded above by the volume of a related cylinder. 

Consider the first set in the intersection, $\{x,y | |g(z)^{\intercal} (x - y) | < \sqrt{3 \alpha(\kappa)} + 2 M R(\kappa)\}$. This set consists of all $x,y$ such that the projection of $x - y$ onto $g(z)$ is less than a vanishing bound, $\frac{1}{\|g(z)\|}(\sqrt{3 \alpha(\kappa)} + 2 M R(\kappa))$. Let $\Omega \times \Omega / g(z)$ denote the quotient space in $\Omega \times \Omega$ perpendicular to $[g(z),-g(z)]$. Then $C(z)$ is contained inside the cylinder defined by requiring that the projection of $[x,y]$ onto $[g(z),-g(z)]$ is less than $\frac{1}{\|g(z)\|}(\sqrt{3 \alpha(\kappa)} + 2 M R(\kappa))$, and that, projecting into the quotient space, $[x,y]$ is within $2 R(\kappa)$ of $z$. 

The volume of the cylinder is proportional to the product of the radii in the quotient space, which is $2T - 1$ dimensional, and the maximum length allowed parallel to $[g(z),-g(z)]$. Therefore:
\begin{equation}
    \text{Vol}(S(\kappa)) \leq \text{Vol}(C(\kappa)) \propto \frac{1}{\|g(z)\|} R(\kappa)^{2T - 1} (\sqrt{3 \alpha(\kappa)} + 2 M R(\kappa)^2)
\end{equation}

Therefore, as long as $g(z) \neq 0$:
\begin{equation}
\text{Vol}(S(\kappa)) = \mathcal{O}_{<=} \left(R(\kappa)^{2T - 1} (\sqrt{3 \alpha(\kappa)} + 2 M R(\kappa)^2) \right).
\end{equation}

It remains to show that $R(\kappa)$ and $\alpha(\kappa)$ can be chosen to converge to zero fast enough to ensure that the probability of sampling $X,Y$ i.i.d.~ from $\pi_x(\kappa)$ in $S(\kappa)$ must go to zero. Both $R(\kappa)$ and $\alpha(\kappa)$ cannot go to zero too quickly. $R(\kappa)$ must go to zero slower than $\kappa$, else there will be a nonzero probability of sampling $X$ outside of $B_{R(\kappa)}(z)$ in the limit. The intermediate bound $\alpha(\kappa)$ cannot go to zero faster than $\kappa^4$, else there is a nonvanishing chance that $\|F_c\|^2 \geq \alpha(\kappa)$. Therefore, we have upper bounds on the fastest possible rates of convergence of each quantity:
\begin{equation}
    R(\kappa) = \mathcal{O}_{>}(\kappa) \text{ and } \alpha(\kappa) = \mathcal{O}_{>}(\kappa^4).
\end{equation}

Suppose we choose $R(\kappa) = \mathcal{O}_{=}(\kappa^a)$, with $a < 1$\footnote{Recall, we also require $a > 4/5$ to ensure that Theorem 1 holds}, and $\alpha(\kappa) = \mathcal{O}(\kappa^b)$ with $b < 4$. The sum $(\sqrt{3 \alpha(\kappa)} + 2 M R(\kappa)^2)$ will be dominated by the slower decaying term in the limit. Therefore, the sum is $\mathcal{O}_{=}(\text{min}(2a,b/2))$. Then:
\begin{equation}
    \text{Vol}(S(\kappa)) \leq \mathcal{O}_{=}(\kappa^d)
\end{equation}
where:
\begin{equation}
    d = (2T - 1)a + \min\{2 a, \frac{1}{2} b\} \text{ given } a \in (4/5,1), b \in (0, 4).
\end{equation}

Our goal is to pick $a$ and $b$ to maximize $d$. Since $a$ can be chosen arbitrarily close to 1, and $b$ can be chosen arbitrarily close to 4:
\begin{equation}
    d < 2 T - 1 + \min{2,2} = 2 T + 1.
\end{equation}
and where $d$ converges to $2T + 1$ as $a$ goes to 1 and $b$ goes to $4$. Of course, the downside of choosing $a$ as close to 1 as possible, and $b$ as close to 4 as possible, is that the error terms associated with the tail of the distribution outside the collapsing balls will vanish slowly, and the probability that $\|F_c\|^2 > \alpha(\kappa)$ will also vanish slowly. The closer each power is to its bound, the slower the tail probabilities converge. Nevertheless, as long as the inequalities $a < 1$, $b < 4$ hold, convergence occurs. A more quantitative theory would optimize these trade-offs. All we aim for here is convergence. 

Thus, we can pick $R(\kappa)$ and $\alpha(\kappa)$ such that $\text{Vol}(S(\kappa))$ vanishes at rate $\mathcal{O}_{=}(\kappa^d)$ for $d$ arbitrarily close to $2T + 1$. 

One complication remains. The sequence of sets $S(\kappa)$ are subsets of the product space $\Omega \times \Omega$, not the original space, $\Omega$. Our concentration definition sets a lower bound on the rate the distribution $\pi_x$ concentrates using nested sets in $\Omega$. 

No matter. Let $S_x(\kappa)$ denote the set of all $x$ for which there is a $y$ such that $x,y \in S(\kappa)$ (the projection of $S(\kappa)$ onto the traits of the first competitor). Define $S_y$ similarly. Then, consider the probability of sampling $X$ and $Y$ in $S(\kappa)$:
\begin{equation}
    \text{Pr}\{X,Y \in S(\kappa)\} = \int_{x \in S_x(\kappa)} \text{Pr}\{Y \text{ s.t. } x,Y \in S(\kappa)\} \pi_x(x,\kappa) dx. 
\end{equation}

Now, $S_x(\kappa)$ are nested, so the domain of integration only shrinks with increasing $\kappa$. Therefore:
\begin{equation}
\begin{aligned}
    \text{Pr}\{X,Y \in S(\kappa)\} & \leq \int_{x \in S_x(\kappa')} \text{Pr}\{Y \text{ s.t. } x,Y \in S(\kappa)\} \pi_x(x,\kappa) dx \leq \left(\sup_{x \in S_x(\kappa')} \text{Pr}\{Y \text{ s.t. } x,Y \in S(\kappa)\} \right) \text{Pr}\{x \in S(\kappa')\} \\
     & \leq \sup_{x \in S_x(\kappa')} \text{Pr}\{Y \text{ s.t. } x,Y \in S(\kappa)\} 
    \end{aligned}
\end{equation}
for $\kappa' \geq \kappa$. 

The set $\{Y \text{ s.t. } x,Y \in S(\kappa)\}$ contains all $y$ within $R(\kappa)$ of $z$, and such that $g(z)^{\intercal} (x - y)$ is less than a vanishing upper bound fixed by $R(\kappa)$ and $\alpha(\kappa)$. The volume of this set is, again, bounded by the volume of a cylinder whose volume is proportional to:
\begin{equation}
    R(\kappa)^{T - 1} (\sqrt{3 \alpha} + 2 M R(\kappa)^2).
\end{equation}

Thus:
\begin{equation}
    \text{Vol}(\{Y \text{ s.t. } x,Y \in S(\kappa)\}) \leq \mathcal{O}_{=}(\kappa^d)
\end{equation}
where now,
\begin{equation}
    d = T - 1 + \text{min}(2 a, \frac{1}{2} b) < T + 1.
\end{equation}

Moreover, $d$ can be chosen arbitrarily close to $T + 1$ by picking $a$ and $b$ arbitrarily close to 1 and 4. In fact, $d > T$ if $2 a > 1$ and $\frac{1}{2} b > 1$, or $a > 1/2$, which was already enforced by requiring $a > 4/5$, and $b > 2$. 

Therefore, provided $R(\kappa) = \mathcal{O}_{<} (\kappa^{1/2})$ and $\alpha(\kappa) = \mathcal{O}_{<}(\kappa^2)$, the volume of $\{Y \text{ s.t. } x,Y \in S(\kappa)\}$ vanishes $\mathcal{O}_{<}(\kappa^T)$ for all $x$. Then, fixing $\kappa'$, the supremum over the volumes must also vanish $\mathcal{O}_{<}(\kappa^T)$ as $\kappa$ goes to zero. Thus, by using the lower bound on the rate of concentration:
\begin{equation}
    \sup_{x \in S_x(\kappa')} \text{Pr}\{Y \text{ s.t. } x,Y \in S(\kappa)\} \rightarrow 0 
\end{equation}
as $\kappa$ goes to zero, so:
\begin{equation} 
    \lim_{\kappa \rightarrow 0 } \text{Pr}\{X,Y \in S(\kappa)\}  = 0. 
\end{equation}

It follows that, as $\kappa$ goes to zero, $\|f(W,Y)\|^2 \geq 3 \alpha(\kappa)$ converges to 1. Therefore:

\vspace{2mm}
\noindent \textbf{Theorem 3: [Almost Sure Transitivity]}
\textit{Assume that the conditions of Theorem 1 hold, and $\pi_x(\kappa)$ concentrates to $z$ at rate $\mathcal{O}_{=}(\kappa)$. Then, if $\nabla_x f(x,y)|_{x=y=z} \neq 0$:}
\begin{equation}
    \lim_{\kappa \rightarrow 0} \text{Pr}\{\mathcal{G}(X,f,\mathcal{E} \text{ is transitive}\} = 1.
\end{equation}

\vspace{2mm}
\noindent \textbf{Proof: } Pick $R(\kappa)$ converging to zero with rate at least $4/5$, and $\alpha(\kappa)$ converging to zero at rate between 2 and 4, say $\alpha(\kappa) = \kappa^3$. Then, under the assumptions of Theorem 1, and given $\nabla_x f(x,y)|_{x=y=z} \neq 0$, Corollary 1.1 applies. Thus, the first condition in Lemma 4 holds. The second follows from the argument given above given that $\pi_x(\kappa)$ concentrates to $z$ at $\mathcal{O}(\kappa)$, and with $\alpha(\kappa) = 3$. Thus, the second condition in Lemma 4 holds, so the competitive network is transitive almost surely in the concentration limit. $\square$

Theorem 3 requires that $\pi_x(\kappa)$ concentrates to $z$ at $\mathcal{O}(\kappa)$, and has covariance $\mathcal{O}_{=}(\kappa^2)$. These conditions are not hard to satisfy simultaneously. For example, any spatially contracting distribution will satisfy all of these conditions. 

\vspace{2mm}
\noindent \textbf{Corollary 3.1:} \textit{If $f$ is nonconstant on open sets, satisfies the smoothness assumptions of Theorem 1, and $\pi_x(\kappa)$ concentrates to $z$ at $\mathcal{O}(\kappa)$ with covariance $\mathcal{O}_{=}(\kappa^2)$, then, for almost all $z$:}
\begin{equation}
    \lim_{\kappa \rightarrow 0} \text{Pr}\{\mathcal{G}(X,f,\mathcal{E} \text{ is transitive}\} = 1.
\end{equation}
\vspace{2mm}

Generic $f$ are nonconstant on open sets, so we expect that, for almost all $z$, there is distribution sufficiently concentrated about $z$, such that any competitive network sampled from the distribution is almost surely transitive.

\section*{Numerical Supplement}
\subsection*{Numerical Model and Bimatrix Games}

To test our theory, we simulate a Gaussian adaptive process on a series of bimatrix games and random performance functions. We test whether evolution promotes concentration towards a small subset of the strategy space, and, consequently, promotes transitivity at the convergence rates predicted by Theorem 2.

Next, we convert the bimatrix payouts into a performance function. Performance could be defined by payout matrices, as when considering individual interactions. However, since our focus is on the population-level dynamics of the network, we consider a population-level payout instead. Note that, expected payout given a pair of mixed strategies is necessarily quadratic. A mixed distribution over two choices is parameterized by one degree of freedom. All quadratic games in one trait are perfectly transitive (see Lemma 1), so would not allow any exploration of convergence to transitivity. In contrast, population level processes allow nontrivial structure. 

For each bimatrix game, we determine event outcomes via a Moran process \cite{lieberman}. We initialize a population of individuals where half adopt strategy $A$ and half adopt $B$. Fitness is determined by game payouts. The process terminates at fixation, i.e.~when all individuals are of one type. The fixation probability given a pair of strategies, acts as a performance function. In this context the game is zero-sum; only one population can fix. Nevertheless, population-level processes can reward cooperation, even in a zero-sum setting, since cooperative agents receive large payouts in predominantly cooperative populations.

\begin{figure}[h!]
    \centering
    \includegraphics[scale=0.4]{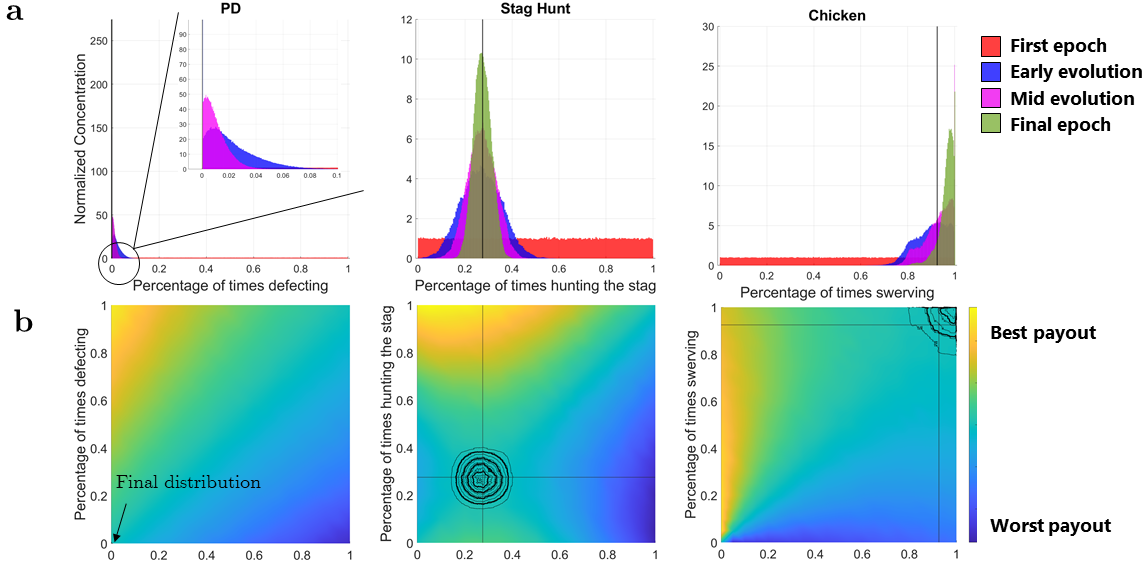}
    \caption{Trait concentration and bimatrix game performances. \textbf{(a)}: The normalized concentration of the strategies for different bimatrix games. The vertical black line is the predicted Nash equilibrium of our performance function. \textbf{(b)}: The performance heatmap for our Moran process. The contours indicate the final locations of the competitors, and the horizontal and vertical lines are the predicted Nash equilibrium.}
    \label{fig:trait distribution and heatmaps}
\end{figure}

Figure \ref{fig:trait distribution and heatmaps} shows the distribution of agents in the trait space at different epochs. In PD, evolution promotes quick convergence to the pure-strategy Nash equilibrium: always cooperate. After the first steps of evolution, nearly all the competitors cooperate in most games (concentrate at $x = 0$). For Stag, the final population distribution resembles a bell curve centered about probability 0.27 to hunt the stag (the NE for the Moran process). As the number of epochs increases, the standard deviation of the trait distribution decreases while the centroid remains constant. Thus, the trait distribution concentrates as time progresses. Likewise, chicken has a truncated Gaussian structure centered near the NE, with standard deviation decreasing over time. Here the distribution abuts the boundary so, projection onto the boundary produces a second mode where agents swerve with probability 1.

Running the Moran process to simulate every event outcome proved prohibitively expensive. Therefore, we approximated the fixation probabilities by interpolating sampled outcomes on a grid. We used the MATLAB curve fitting tool to generate a performance function that closely approximates the fixation probabilities. For the stag hunt and prisoner's dilemma games, cubic polynomials fit the data with no visible systematic errors (R-squared values $\geq$ 0.9993). We used the fits to compute the gradients and Hessian used throughout the convergence theory. For chicken, the extreme behavior at the corners of the strategy space (see the cusp in the bottom left hand corner of Figure \ref{fig:trait distribution and heatmaps}) created systematic errors so we used a cubic spline interpolant instead, and used numerical differentiation to approximate the gradient and Hessian.

Competitive outcomes between any pair of competitors were then determined using the performance function. Each competitor plays some number of games (usually between 100 and 1000) against a random selection of the other competitors drawn from the current population. We then selected competitors to reproduce with a top-10\% cutoff model, a Softmax selection criteria, or a logit criteria. In the top 10\% cutoff model, the 10 percent of competitors with the highest win rate are selected to reproduce, and each produce 10 children. Children are assigned the same traits as their parent, plus a normal random vector scaled by a genetic drift parameter. If the sampled traits fall outside the trait space, they are projected back onto the boundary of the trait space instead. In the softmax model, for each child, a competitor's probability of reproducing was a function of their fitness compared to the average fitness, with a softmax exponential (default 5) applied.  Every phase of competition, selection, and reproduction is one ``epoch".

At every epoch in the process, we perform a Helmholtz-Hodge Decomposition (HHD) of the complete graph of competitors to calculate the size of the transitive and intransitive components. These are normalized to produce proportions of transitivity and cyclicity as defined in \cite{strang2020applications,strang2022network}. To evaluate concentration, we calculated the covariance of the traits and the number of clusters, counted using a Gaussian mixture model. We allowed evolution to proceed until epoch 50, or when we had one stable cluster, whichever came first.

As a sanity check, we compared the location of the final trait distributions, to the Nash equilibria for the bimatrix games. Prisoner's dilemma has two pure-strategy Nash equilibria: either both agents always cooperate or always defect. The always cooperate strategy, has a higher expected payout for both competitors, but is invasible by a defecting strategy. In stag hunt, there are two pure-strategy Nash equilibria along with a mixed-strategy to hunt the stag with probability 0.25. For chicken, there is a pure-strategy Nash equilibrium to always go straight, and a mixed-strategy equilibrium to swerve with probability 0.999. While these Nash equilibria provide approximate locations for evolution to settle, in practice, the payout governing evolution is directed by the fixation properties from the Moran process. These equilibria are close to, but not exactly, the equilibria observed for the bimatrix games. In general our processes converged towards stable distributions centered near the Nash equilibria of the Moran process (see Figure \ref{fig:trait distribution and heatmaps}).

In most experiments ran we observed convergence towards transitivity driven by increasing concentration. Our theory predicts the rate of convergence to transitivity in concentration. This rate depends on the smoothness of the performance function, as measured by the norms of its low order partial derivatives. We tested the accuracy of these predicted rates as follows. We used the Trait Performance Theorem from \cite{cebra2023similarity} by comparing the average sizes of the transitive and cyclic components in randomly sampled ensembles about the final cluster centroid. We drew each ensemble from a normal distribution, centered at the final cluster centroid, while taking the covariance to zero. We compared our empirical estimate to our analytic prediction based on $\epsilon$ (see \cite{cebra2023similarity}) to confirm the rates of convergence predicted in Table 1. 

In Table \ref{table: bimatrix parameters}, we see the default parameters for the bimatrix games simulations in the paper.

\begin{table}[h]
\begin{centering}
\begin{tabular}{ |p{4cm}| p{4cm}|p{4cm}|  }
 \hline
 \multicolumn{3}{|c|}{List of variables of consideration} \\
 \hline
 Parameter& Control & Perturbations \\
 \hline
 Fit (PD and stag) &   cubic  & quintic\\
 Interpolant (Chicken) & cubic spline & \\
 Genetic drift& $5 \cdot 10^{-4}$  & $5 \cdot 10^{-5}$, $1 \cdot 10^{-4}$, $5 \cdot 10^{-3}$, $1 \cdot 10^{-3}$, $1 \cdot 10^{-2}$, $5 \cdot 10^{-2}$\\
 Games played per epoch & 100  & 10, 1000\\
 Softmax power & 5 & 2\\
 \hline
\end{tabular}

\end{centering}
\vspace{4mm}
 \caption{List of parameter values and variations tested in the bimatrix games.}
 \label{table: bimatrix parameters}
\end{table}

While the bimatrix games provide a widely studied measure for competition, they all have a one-dimensional trait space. Games in one-dimensional spaces are a special, highly transitive case (see Lemma 1). Therefore, we also considered randomly-generated, $n$-dimensional performance functions with tuneable structure chosen to illustrate the generality of our theory.

\subsection*{Random Performance Functions}

We designed our performance functions to satisfy the following properties. First, we desired tuneable smoothness, so we constructed the function as a sparse sum of Fourier modes with variable amplitudes and frequency. By increasing the low order modes we promote transitivity over neighborhoods of a fixed size. Second, we ensured that distinct traits of distinct competitors interact to produce attribute tradeoffs (e.g.~ speed versus strength). Such tradeoffs are, generically, the source of cyclic competition arising at lowest order in the Taylor expansion of performance. In special cases when distinct traits do not interact, the degree of cyclic competition vanishes exceptionally fast during concentration. Lastly, we skew symmetrized our function to enforce fairness, that is, $f(x,y) = -f(y,x)$.

The resulting performance function took the form:
\begin{equation}
\begin{aligned}
    f(x,y|\mathcal{P},\alpha,\phi,m) = \sum_{k=1}^m \sum_{i,j \in \mathcal{P}(k)} \frac{\alpha_{i,j}(k)}{k^2}  & \left( \sin(2 \pi k (x(i) - \phi_{i,j}(k))) \cos(2 \pi k (y(j) - \phi_{i,j}(k))) - \hdots \right. \\
    & \left. \sin(2 \pi k (y(i) - \phi_{i,j}(k))) \cos(2 \pi k (x(j) - \phi_{i,j}(k))) \right)
\end{aligned}
\end{equation}
where $\mathcal{P}(k)$ is the collection of interacting trait pairs at frequency $k$, $\alpha$ are the amplitudes, $\phi$ are a collection of phase shifts, and $m$ is the max frequency (and number of distinct frequencies) considered. Note that taking a difference of the form $f(x) g(y) - f(y) g(x)$ is automatically alternating in $x$ and $y$. We divided the amplitudes at higher frequencies by $k^2$ so that each frequency contributes equally to the Hessian. 

Then, our evolution test continued in the same manner as before, initialized with a uniformly sampled population selected from an $n$-dimensional trait hypercube with sides [-1,1].

\begin{table}[h]
\begin{center}
\begin{tabular}{ |p{4cm}| p{4cm}|p{4cm}|  }
 \hline
 \multicolumn{3}{|c|}{List of variables of consideration} \\
 \hline
 Parameter& Control Value & Perturbations \\
 \hline
Number of competitors   & 250    & 50, 100\\
 Number of traits&   4  & 2, 8, 16, 32\\
 Number of trig. modes &2 & 4, 6\\
 Trig. mode amplitude    &1 & 0.5\\
 Linear mode amplitude&   1  & 0, 0.5\\
 Genetic drift& $5 \cdot 10^{-4}$  & $1 \cdot 10^{-3}$, $5 \cdot 10^{-3}$, $1 \cdot 10^{-4}$\\
 Games per competitor& 100  & 1, 5, 10\\
 Softmax power& 5 & 2\\
 \hline
\end{tabular}
\end{center}
\vspace{2mm}
 \caption{A list of parameters considered in the random performance functions example.}
 \label{table: random performance functions parameters}
\end{table}

In Table \ref{table: random performance functions parameters} we see the control and tested parameters that we used for the random performance functions.

\begin{figure}[t]
    \centering
    \includegraphics[scale=0.6]{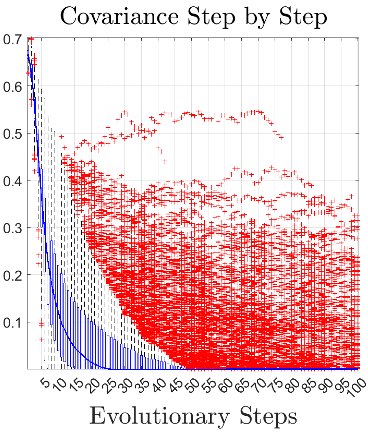}
    \caption{Calculated Kendall intransitivity by evolutionary epoch for the random performance functions test.}
    \label{fig:kendall intransitivity}
\end{figure}

To confirm the concentration assumptions, we tracked the covariance of the competitor traits by evolutionary step as well. In Figure \ref{fig:kendall intransitivity} we can observe that the covariance shrinks towards 0 as evolution proceeds, confirming that our evolutionary mechanism promotes competitor concentration.

\subsubsection*{Genetic Drift Test}
\begin{figure}[h!]
    \centering
    \includegraphics[scale=0.303]{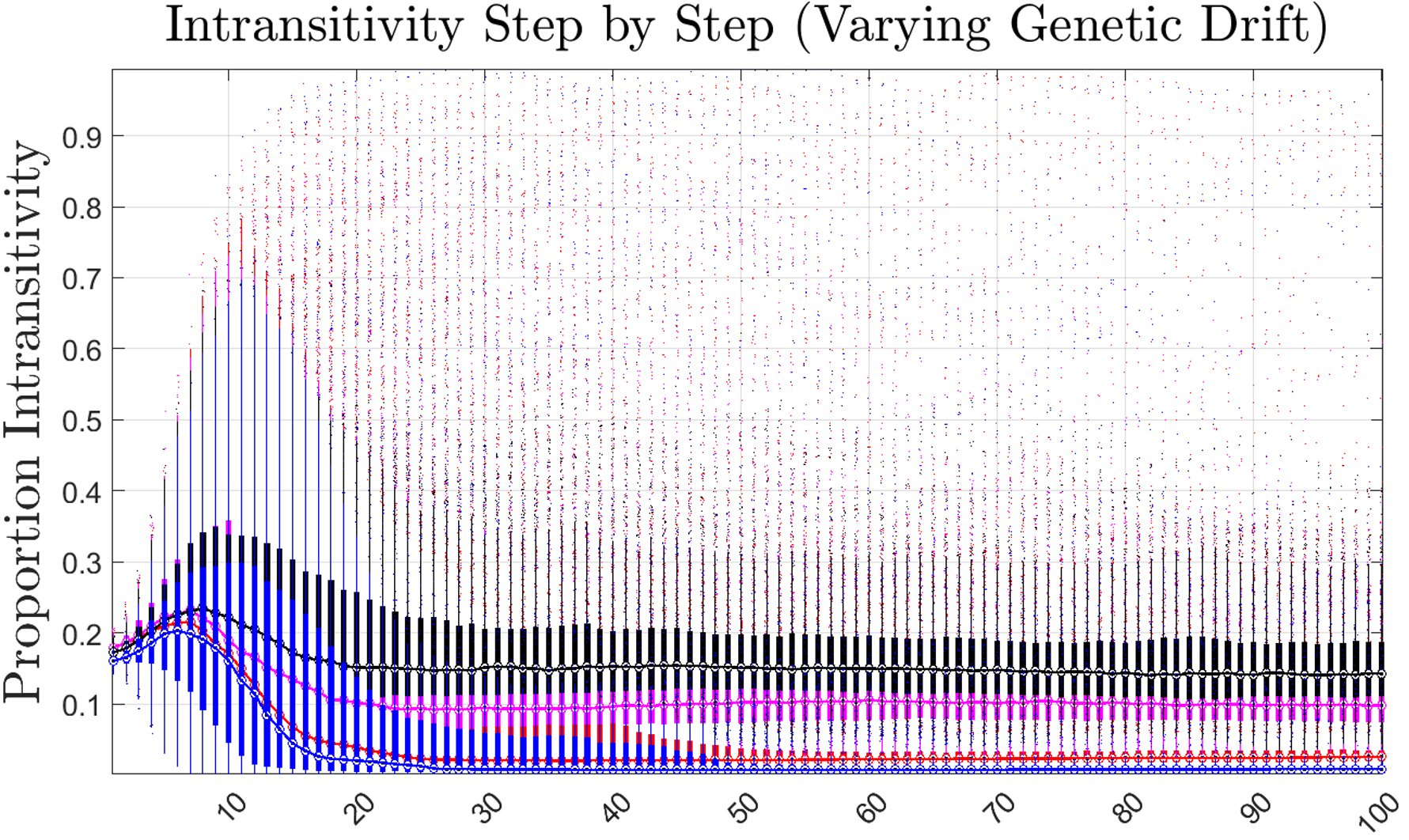}
    \caption{Step-by-step intransitivity for random performance functions. Colors corresponding to different genetic drifts: \textbf{(Blue):} $5 \cdot 10^{-4}$, \textbf{(Red):} $1 \cdot 10^{-3}$ (control), \textbf{(Magenta):} $5 \cdot 10^{-3}$, \textbf{(Black):} $1 \cdot 10^{-2}$.}
    \label{fig:genetic drift intransitivity step-by-step}
\end{figure}

In this case, changing the genetic drift does affect whether the system converged to transitivity. Figure \ref{fig:genetic drift intransitivity step-by-step} shows the results. Increasing the genetic drift by an order of magnitude above the control values yielded an increase in the steady-state intransitivity, largely because the increased genetic drift prevents the competitors from concentrating as in our assumptions.

\end{document}